\documentclass[]{interact}

\usepackage{epstopdf}
\usepackage[caption=false]{subfig}
\usepackage{url}
\usepackage{hyperref}
\usepackage{cleveref}
\usepackage{enumitem}
\usepackage[authoryear]{natbib}
\usepackage{dsfont}
\usepackage{float}

\bibpunct[, ]{[}{]}{,}{n}{,}{,}

\theoremstyle{plain}

\theoremstyle{definition}

\theoremstyle{remark}

\begin{document}

\title{Structural Under-Representation of Women in News: Nonparametric Bayesian Mixtures Capture Time-Dependent Dynamics}

\author{
\name{Isabella Habereder\textsuperscript{a,b}\thanks{CONTACT Isabella Habereder. Email: i.habereder@media-bias-research.org}, Thomas Kneib\textsuperscript{a}, Isao Echizen\textsuperscript{b} and Timo Spinde\textsuperscript{b}}
\affil{\textsuperscript{a}Chair of Statistics and
Econometrics, Georg-August-University Göttingen,
Humboldtallee 3, 37073 Göttingen, Germany; \textsuperscript{b}Information and Society Research Division, National Institute of Informatics (NII), 2-chome-1-2 Hitotsubashi, Chiyoda City, Tokyo 101-8430, Japan}
}

\maketitle

\begin{abstract}
The under-representation of women as sources cited in news media is one prominent representation of gender bias. Understanding where gender bias concentrates and how it evolves is essential for targeted mitigation.
Because gender representation varies across topics, time, and reported-on regions, creating complex dependencies that are difficult to capture parametrically, we employ a nonparametric model to uncover latent cluster structures and temporal dynamics. 
We combine time-dependent Bayesian mixture modeling techniques with a Beta mixture kernel tailored to female quote shares, bounded between 0 and 1. 
Fitted on Canadian news articles from 2019 to 2024, the model reveals structural under-representation of women across all clusters, with news topic driving differences in female quote shares more strongly than the reported-on region. More than 85\% of topic-region time series show no improvement toward gender parity over the observation period. Dynamic density estimation confirms that the aggregate distribution of female quote shares remains stable between 2019 and 2024.
Our application demonstrates that advanced probabilistic models not only reproduce findings in gender bias research but also reveal latent dependencies and structural patterns that simpler approaches miss, encouraging future adoption of model-based frameworks for studying media bias.

\end{abstract}


\section{Introduction}\label{sec:introduction}
The under- and misrepresentation of one gender in the media relative to other genders is referred to as gender bias \cite{rao2021gender, riedl2024journalistic}.
Research on gender bias indicates that the under- and misrepresentation of women in media contribute to the construction of gender roles by providing a foundation for stereotypical beliefs \cite{costa_jussa2019gender, power2019women}. Gender roles and stereotypes contribute to the perpetuation of structural social inequalities by reinforcing categorical distinctions and shaping individual identity and position within established social hierarchies \cite{dacon2021gender}. Furthermore, gender bias not only influences society but is also, in turn, influenced by social structures \cite{hooghe2015enduring, riedl2024journalistic, shor2015paper, vanderpas2020gender}, such as the dominance of men in leadership positions across fields like politics, business, and sports \cite{shor2015paper}.

Existing approaches to identifying and quantifying observable patterns of gender bias in media largely rely on manual content analysis \cite{joshi2020violators, lee2024visuality, riedl2024journalistic, trimble2021gender}, descriptive statistics \cite{taboada2025reported}, or computational approaches that use predefined lexicons, topic models, or keyword frequencies to capture gendered language and thematic differences  \cite{dacon2021gender, jia2016women, rao2021gender}. Some studies also employ regression-based frameworks \cite{hooghe2015enduring, shor2015paper, shor2019women} to estimate the association between gender bias and selected covariates while controlling for confounding factors.
While these methods have provided important insights, they typically depend on strong parametric assumptions, limiting their ability to capture the complex and latent structures in gender bias data. 
For example, gender bias often exhibits temporal dependencies \cite{deiorio2023bayesian}, and identifying the driving factors behind gender bias is difficult \cite{hooghe2015enduring, riedl2024journalistic, shor2019women}, since, as stated above, gender bias is mutually influenced by social structures.

Therefore, we require a methodology that is sufficiently general and flexible yet computationally simple.
To address the conceptual and methodological challenges outlined above, we propose a Bayesian nonparametric model for analyzing gender bias data. One of the most used models in the class of Bayesian nonparametric models is the Dirichlet process (DP) mixture model \cite{gutierrez2016airquality}. In the DP mixture setting, one assumes observations $y$ come from the mixture density, which is given by $f(\cdot, G) = \int_{\Theta}k(\cdot,\theta)dG(\theta)$, where $\Theta$ is the parameter space and $k$ is the mixture kernel. The mixture distribution $G$ is assigned a DP prior, which admits the representation \cite{sethuraman1994constructive}
$G = \sum_{h=1}^{\infty}w_{h}\delta_{\theta_h}$, 
where $\delta_{x}$ is the point measure centered on a fixed point $x$. The weights $w_{1}, w_{2}, \dots$ follow the construction specification $w_1 = v_1$, $w_h = v_h \prod_{l < h} (1-v_l)$ with $v_h \overset{\text{iid}}{\sim}\text{Beta}(1,M), \; M > 0$, independent of $\theta_h$.

In some applications \cite{deiorio2023bayesian, gutierrez2016airquality}, it is of interest to study changes in the distribution of the observation $y$ as a function of time $t$, i.e., we need a dynamic version of the model. Due to the applications described in \cref{sec:application}, we will focus on the discrete time case, i.e., when $t \in \{1, 2, \dots, T \}, \; T < \infty$. The mixture density in the dynamic setting is given by $f_t(\cdot, G) = \int_{\Theta}k(\cdot,\theta)dG_t(\theta)$. \citet{deiorio2023bayesian} proposed for the purpose of dynamic clustering and density estimation of gender stereotype data a Bayesian nonparametric auto-regressive (order 1) Dirichlet process mixture model (AR1-DP model), where the time-dependent mixture distribution $G_t$ admits the representation
$G_t = \sum_{h=1}^{\infty}w_{t,h}\delta_{\theta_h}$.
The weights $w_t = (w_{t,1}, w_{t,2}, \dots)^T$ follow the construction specification
$w_{t,1} = \xi_{t,1}$, and $w_{t,h} = \xi_{t,h}\prod_{l=1}^{h-1}(1-\xi_{t,l})$
with
$\xi_{t,h} = 1- (1-\Phi(\epsilon_{t,h}))^{\frac{1}{M}}$,
where $\Phi$ denotes the cumulative distribution function of the standard normal distribution.
The latent process $(\epsilon_{t,h})_{t \geq 1}$ follows an auto-regressive process of order 1, given by
$\epsilon_{1,h} \overset{\text{iid}}{\sim} \mathcal{N}(0,1)$, and $\epsilon_{t,h}\vert \epsilon_{t-1,h}, \psi \overset{\text{iid}}{\sim} \mathcal{N}(\psi \epsilon_{t-1,h}, 1- \psi^2)$, where $\psi \in (-1,1)$. \citet{deiorio2023bayesian} decided on a Gaussian kernel for the choice of the mixture kernel.

Since a Gaussian kernel does not respect a bounded support of the response $y$, we adapt the AR1-DP model for our use case by using a Beta-mixture kernel, which, to the best of our knowledge (see discussion in \cref{sec:related_work}), is the first application of such a kernel in a dynamic setting. This extension enables us to investigate clustering dynamics in time-dependent gender bias data and to perform flexible density estimation, thereby making the underlying probabilistic structure of the data explicit and analytically accessible. Furthermore, our approach provides a general framework for studying temporal patterns of bias in media data, allowing for the analysis of under-representation dynamics over time.

\section{Related Work}\label{sec:related_work}
As already stated in the \cref{sec:introduction}, the AR1-DP model of \citet{deiorio2023bayesian} accommodates both dynamic clustering and dynamic density estimation; we briefly outline alternatives for each approach.
Regarding dynamic clustering, \citet{page2022dependent} offers an alternative to the approach of \citet{deiorio2023bayesian}. Instead of first modeling dependent random measures and then inducing partitions from them, \citet{page2022dependent} models the sequence of partitions directly. However, the temporal random partition model (tRPM) in \cite{page2022dependent} is purely a partition model and is therefore not specifically designed for dynamic density estimation. On the other hand, \citet{gutierrez2016airquality} offers an alternative when the focus is on dynamic density estimation, but does not provide dynamic clustering.

The non-parametric Bayesian mixture model of \citet{deiorio2023bayesian} is constructed with a so-called copula-based time-dependent DP.
Because of the challenging posterior distribution (e.g., through its hierarchical integration into the DP process) of the copula-based construction of the DP, \citet{lee2025logistic} proposes a more computationally efficient logistic-Beta process as an alternative for the copula-based construction. 
Moreover, \citet{marin2026bayesian} integrates the ideas of \citet{lee2025logistic} into the model of \citet{deiorio2023bayesian} and adopts a Stirling-Gamma prior to ease the process of incorporating prior knowledge. But, the computational efficiency of the innovations introduced, e.g., by \citet{lee2025logistic}, is negligible for our model (see \cref{sec:methodology}), and we prefer the copula-based approach for its flexibility.

The AR1-DP model of \citet{deiorio2023bayesian}, as well as related models (e.g., \cite{castro2021measuring, griffin2011stick, griffin2006order, lee2025logistic}) employ a Gaussian kernel as a choice for $k$, primarily motivated by computational convenience: pairing a Gaussian kernel with a Normal-Gamma base distribution yields conjugate full conditionals, thereby avoiding Metropolis-Hastings steps for the cluster parameters. While mixtures of Gaussian distributions are dense in the space of continuous densities on $\mathbb{R}$ \cite{ferguson1973bayesian, lo1984class}, this choice constitutes a formal support misspecification when observations are bounded. \citet{gutierrez2016airquality} partially addresses this by adopting a log-Normal kernel, which respects the positive half-line at the cost of requiring a logarithmic transformation. For our application, the observations are proportions taking values in $(0, 1)$, for which a Beta distribution as the mixture kernel is the natural choice, since it respects the support. The absence of a conjugate base distribution for the Beta kernel necessitates Metropolis-Hastings sampling for the cluster parameters (see \cref{sec:methodology}). 
For static (i.e., time-independent) DP mixture models, there is already research on using the Beta distribution as a mixture kernel. \citet{kottas2006dirichlet} propose a class of non-parametric Bayesian DP mixture models with a Beta distribution providing the kernel. 
Furthermore, \citet{kottas2007spatial} proposes a bi-variate Beta distribution as the mixture kernel for a DP model, motivated by the analysis of spatial point patterns in a bounded region. However, to the best of our knowledge, a dependent Beta DP mixture model for dynamic clustering and density estimation has not appeared in the literature.

\section{Gender Bias Dataset}\label{sec:data}
To illustrate our methodology, we have chosen a dataset that has already yielded promising results in related work \cite{rao2021gender} on the topic-dependence of gender bias. We use data obtained by the Gender Gap Tracker\footnote{\url{https://gendergaptracker.informedopinions.org/}; The Gender Gap Tracker is a joint project between Informed Opinions, a non-profit organization focused on amplifying women's voices in media, and Simon Fraser University, facilitated through the Discourse Processing Lab and the Big Data Initiative.}. The dataset comprises text and metadata from daily news articles published by Canadian news outlets since October 2018. We have approximately 3.5 million data points available for analysis. Among others, the datasets' attributes are: \textit{title}, \textit{text corpus}, \textit{date of publication}, \textit{number of female quotes}, and \textit{number of male quotes}\footnote{We slightly renamed the original attributes.}. Due to the dataset's restriction to female- and male-annotated quotes, we also employ a binary classification scheme for gender in our project. 

Gender bias in the news can be measured in different ways, such as by the number of female quotes in an article \cite{rao2021gender, taboada2025reported}, the allotted speaking time of women in television news \cite{hooghe2015enduring}, the number of female-related names mentioned \cite{shor2019large}, or through the distribution of gendered wording \cite{dacon2021gender, shor2019women, trimble2021gender}. Due to the data, we calculate the share of female quotes in a news article $a$, thereby providing a measure of the coverage of female voices in the news landscape:
    \begin{equation}\label{eqn:gender-bias-measurement-article}
        \text{GenderBias}(a) = \frac{\text{number of female quotes in } a}{\text{total number of quotes in } a}.
    \end{equation}
We processed and extended the data as described in detail in \cref{appx:data-preprocess} in the appendix. We extended the data by the article's topic and the gender bias measurement (see equation \eqref{eqn:gender-bias-measurement-article}). There is evidence that the extent of gender inequality in representation depends on the cultural region where the reporting is about \cite{elmasry2025gender, greenwald2023israeli, joshi2020violators, lee2024visuality}, therefore we additionally extended the attributes to include the narrative location (i.e., the country the article reports on) and the narrative region (i.e., the cultural region \cite{kolb1962, newig2014} the article reports on). We performed two different aggregation levels. First, we aggregated the data at a country $\times$ topic $\times$ year level ($N=9\,534$), and second, at a region $\times$ topic $\times$ year level ($N=990$). A list of all included countries and cultural regions can be found in \cref{tab:list-of-regions-and-countries} in the appendix.
We denote the aggregated data point at each timestep in the following with \textit{observation}.
The value of our target variable, i.e., the gender bias measurement for observation $i$ at time $t$ is obtained by using \eqref{eqn:gender-bias-measurement-article} and is defined as
    \begin{equation}\label{eqn:gender-bias-measurement-observation}
        y_{i,t} = \frac{\sum_{a \in N_{i,t}}\text{GenderBias}(a)}{\# N_{i,t}},
    \end{equation}
where $N_{i,t}$ denotes the set of articles included in the aggregation for observation $i$ at time $t$.



\section{Methodology: A Bayesian Non-parametric Temporal Dynamic Clustering Approach via Auto-regressive Dirichlet Priors}\label{sec:methodology}
Let $M > 0$, $\psi \in (-1,1)$ be fixed and $G_0$ denotes a Normal-Gamma base distribution. The process $\text{AR1-DP}(M, \psi, G_0)$ is defined by the following construction:
Let the atom parameters be defined as 
$\theta_h \overset{\text{iid}}{\sim} G_0$ for $h \geq 1$.
We set 
$G_t = \sum_{h=1}^{\infty}w_{t,h}\delta_{\theta_h}$ 
where $\delta_{x}$ is the point measure centered on a fixed point $x$. The weights $w_t = (w_{t,1}, w_{t,2}, \dots)^T$ follow the construction specification
$w_{t,1} = \xi_{t,1}$, and $w_{t,h} = \xi_{t,h}\prod_{l=1}^{h-1}(1-\xi_{t,l})$ (for $h > 1$)
with
$\xi_{t,h} = 1- (1-\Phi(\epsilon_{t,h}))^{\frac{1}{M}}$ for $h \geq 1$,
where $\Phi$ denotes the cumulative distribution function of the standard normal distribution.
For $h \geq 1$, the latent process $(\epsilon_{t,h})_{t \geq 1}$ follows an auto-regressive process of order 1, given by
$\epsilon_{1,h} \overset{\text{iid}}{\sim} \mathcal{N}(0,1)$, and $\epsilon_{t,h}\vert \epsilon_{t-1,h}, \psi \overset{\text{iid}}{\sim} \mathcal{N}(\psi \epsilon_{t-1,h}, 1- \psi^2)$  (for $t > 1$).

For our application, we use a finite-truncated version of the process:
Let $Y_{i,t}$ denote a random variable where its realization at time $t=1,\dots, T$ and observation $i=1,\dots, N$ is $y_{i,t}$ from \eqref{eqn:gender-bias-measurement-observation}. We write $y_t = (y_{1,t}, \dots, y_{N,t})^T$ and $y_{1:T} = y_1, \dots, y_T$. 
Furthermore, let the parameter vector for $j=1,\dots, J$ be $\theta_j = (m_j, \phi_j)$ and we write $\theta = (m_{1:J}, \phi_{1:J})$. We also write for the finite-truncated case $w_t = (w_{t,1}, \dots, w_{t,J})^T$. We define our model through
\begin{eqnarray*}
    Y_{i,t} \vert s_{i,t}, \theta_{s_{i,t}} &\overset{\text{ind}}{\sim}& \text{Beta}(\mu_{i,t}\phi_{s_{i,t}},\; (1-\mu_{i,t})\phi_{s_{i,t}}) \; \text{for} \; i =1, \dots, N \\
    \mu_{i,t} &=& \text{logit}^{-1}(m_{s_{i,t}}) \\
    s_{i,t} \vert w_t & \overset{\text{iid}}{\sim} & \sum_{l=1}^{J}w_{t,l}\delta_l \; \text{for} \; i =1, \dots, N \\
    \theta_{s_{i,t}} & \overset{\text{iid}}{\sim} & G_0 \; \text{for} \; s_{i,t} =1, \dots, J\\
    (G_t)_{t \geq 1} &\sim & \text{AR1-DP}(M, \psi, G_0).
\end{eqnarray*}
The goal is to sample from the joint posterior distribution $p(\theta, s_{1:T}, w_{1:T}, \epsilon_{1:T} \vert y_{1:T})$. The following conditional independencies (by construction) are incorporated into the sampling algorithm. (1) $s_t \perp\!\!\!\perp s_{t^\prime} \vert w_t, \theta$ for all $t \neq t^\prime$; (2) $\theta_j \perp\!\!\!\perp \theta_{j^\prime} \vert s_{1:T}$ for all $j \neq j^\prime$; (3) $(w_{1:T}, \epsilon_{1:T}) \perp\!\!\!\perp \theta \vert s_{1:T}$. Therefore, the design of a Gibbs-motivated sampler is straightforward. \\
\textbf{Step 1:} Sample from $p(s_{i,t} = j \vert w_t, \theta, y_{i,t})$ for all $j = 1, \dots, J$; $t = 1, \dots, T$; $i = 1, \dots, N$ by Gibbs sampling. By (1), the allocation of observation $i$ at time $t$ depends only on the current weights $w_t$ and cluster parameters $\theta$, so each $s_{i,t}$ can be drawn independently.\\
\textbf{Step 2:} Sample from $p(\theta_j \vert s_{1:T}, y_{1:T})$ for all $j = 1, \dots, J$ by Metropolis Hastings (MH). By (2), the clusters are conditionally independent given $s_{1:T}$, so each $\theta_j$ can be updated separately using only observations assigned to cluster $j$, pooled across all time points. We employ two separate one-dimensional random-walk MH steps per cluster per iteration. In the first step, $m_j$ is updated conditional on $\phi_j$ via the proposal $\tilde{m}_j = m_j + \zeta, \; \zeta \sim \mathcal{N}(0, \sigma_{m,j}^2)$. In the second step, $\log\phi_j$ is updated conditional on $m_j$ via the proposal $\log\tilde{\phi}_j = \log\phi_j + \zeta, \; \zeta \sim \mathcal{N}(0, \sigma_{\phi,j}^2)$. Each proposal standard deviation $\sigma_{m,j}$ and $\sigma_{\phi,j}$ is tuned independently per cluster via Robbins-Monro stochastic approximation \citep{robbins1951stochastic}, targeting the optimal one-dimensional acceptance rate of $0.44$ \citep{roberts2001optimal}. \\
\textbf{Step 3:} Sample $\epsilon_{1:T}$ and recompute $w_{1:T}$. By (3), $\epsilon_{1:T}$ is conditionally 
independent of $\theta$ given $s_{1:T}$, so the cluster assignments are sufficient to couple the weight process to the observed data. For each $t = 1, \dots, T-1$ and $j = 1, \dots, J-1$, we sample from $p(\epsilon_{t,j} \vert \epsilon_{t,-j}, s_t, \psi, M)$. Since it has no closed form due to the nonlinear construction of $w_t$ through $\epsilon_t$ (see \cref{sec:methodology}), direct Gibbs sampling is infeasible, and we resort to a random walk MH with symmetric Gaussian proposal $\sim \mathcal{N}(\epsilon_{t,j}, \hat{\sigma}^2_\epsilon)$. The proposal standard deviation $\hat{\sigma}_\epsilon$ is analogously to Step 2 tuned adaptively via Robbins-Monro stochastic approximation \cite{robbins1951stochastic} targeting the optimal one-dimensional acceptance rate of $0.44$ \cite{roberts2001optimal}. 


\section{Application: Gender Bias in the Canadian News}\label{sec:application}
We report the key findings from our application at the region $\times$ topic $\times$ year aggregation level. The key findings for the country $\times$ topic $\times$ year aggregation level are described in \cref{appx:country-level} in the appendix. The parameter choices for our AR1-DP model (see \cref{sec:methodology}) are discussed in \cref{appx:paramtuning} of the appendix. We set $M= 0.4$, $\psi = 0.85$ and $J=15$. Prior-hyperparameters of the base distribution $G_0$ were selected using an empirical Bayesian moment-matching procedure (see formulas in \cref{appx:country-level}) based on the pooled observations across all time periods. To improve early-stage mixing, the first five mixture components were re-initialized using empirical quantiles of the pooled data: component means were set to the logit-transformed 10th, 30th, 50th, 70th, and 90th percentiles, while corresponding precision parameters were set using a moment-based estimator based on the global sample variance. The remaining components were left unchanged. Furthermore, we performed $ 50 \, 000$ iterations of the MCMC algorithm described in \cref{sec:methodology}, with a burn-in period of $25 \, 000$. 

\subsection{Cluster Structure}\label{sec:cluster-structure}
For summarizing the samples for the cluster allocations $s_{1:T}$ from the posterior distribution for the region $\times$ topic $\times$ year aggregation level, we use the symmetric binder loss criterion \cite{binder1978bayesian}. For each time point, we compute the posterior co-clustering matrix $\hat{C}_{ij} = \frac{1}{S}\sum_{s=1}^{S}\mathds{1}[s_i^{(s)} = s_j^{(s)}]$ and apply average-linkage hierarchical clustering to the dissimilarity $D_{ij} = 1 - \hat{C}_{ij}$, cutting at threshold $\tau = 0.5$. This minimizes the expected Binder loss under equal misclassification costs \cite{lau2007bayesian}. 

\begin{figure}[ht]
    \centering
    \includegraphics[width=\linewidth]{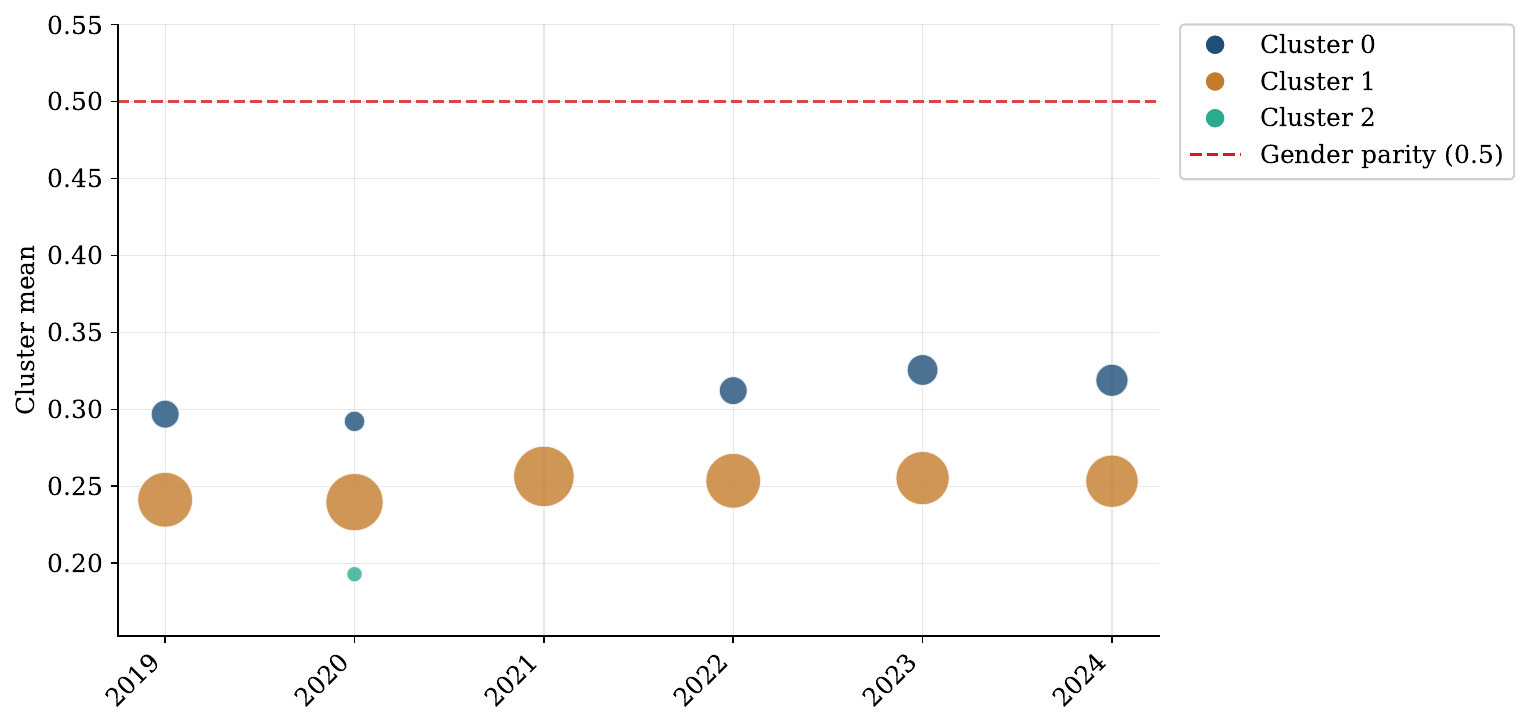}
    \caption{Illustration of the cluster mean over time. The bubble size is proportional to the number of observations.}
    \label{fig:bubble-chart}
\end{figure}

As shown in \cref{fig:bubble-chart}, across all regions, topics, and years covered in this study, the posterior mean female source share falls substantially below gender parity (i.e., mean $< 0.5$). The overall range of cluster means spans 0.19 to 0.31, confirming that the structural under-representation of female voices in Canadian news is not an artifact of any particular regional or thematic focus but a pervasive feature of the data. At the same time, the width of this range ($=0.12$) indicates that the degree of under-representation varies meaningfully and in a structured way. We identify three clusters. One of these clusters dominates the data in terms of size $N$, i.e., the total number of distinct region $\times$ topic combinations over all time points. Cluster 1 ($N = 833$, mean $= 0.25$) together accounts for $84.1\%$ of all observations. The remaining two clusters are considerably smaller; cluster 2 (mean $= 0.19$) contains four observations, and cluster 0 (mean $= 0.31$) contains 153 observations.  

Inspection of the topic composition (see \cref{fig:topic-composition} in \cref{appx:figures} in the appendix) within each cluster reveals a striking pattern: cluster membership is determined by the topic of reporting. Cluster 0, the cluster with the highest female source share (mean $= 0.31$), is constituted by topics 2 (public health/healthcare) and 5 (culture), which together account for $57.5\%$ of its observations (topic 2: $25.5\%$, topic 5: $32.0\%$). Cluster 1, the largest cluster (mean $= 0.25$), shows an approximately uniform topic distribution. The low-end cluster, cluster 2 (mean $= 0.19$), is concentrated in Topics 11 (sport) and 13 (mobility).

By contrast, the distribution of geographic regions across clusters is more uniform (see \cref{fig:region-composition} in \cref{appx:figures} in the appendix). In the dominant cluster (cluster 1), all eleven regions contribute approximately equal shares of observations. In cluster 0, the regions Australia/Oceania and Anglo America have the largest share of observations, even less strikingly. The relative regional uniformity of cluster 0 stands in contrast to the topic pattern, where two topics (topic 2 - public health/healthcare; topic 5 - culture) alone account for over half of cluster 0.   

The region $\times$ topic heatmap (see \cref{fig:region-topic-heatmap} in \cref{appx:figures} in the appendix)  makes the dominance of topic effects visually explicit. Columns for topic 2 (public health/healthcare) and topic 5 (culture) are uniformly warm-colored across all rows, reflecting consistently 'higher' female source shares irrespective of region. Columns for, for example, topics 3 (conflicts), 4 (corporate), and 6 (economy) are uniformly cool, indicating low female-source shares across all regions. Row-level variation, which would indicate regional effects, is visibly smaller than column-level variation throughout the matrix. 
To formally quantify the relative contribution of regional versus temporal variation, we compute, for each topic separately, the between-region variance (variation in region-level means) and the within-region variance (average temporal variability within a region). The variance decomposition (see \cref{fig:variance-analysis} in \cref{appx:figures} in the appendix) supports the findings of the region $\times$ topic heatmap.

\subsection{Temporal Dynamics}\label{sec:temporal_dynamics}
\begin{figure}[ht]
    \centering
    \includegraphics[width=\linewidth]{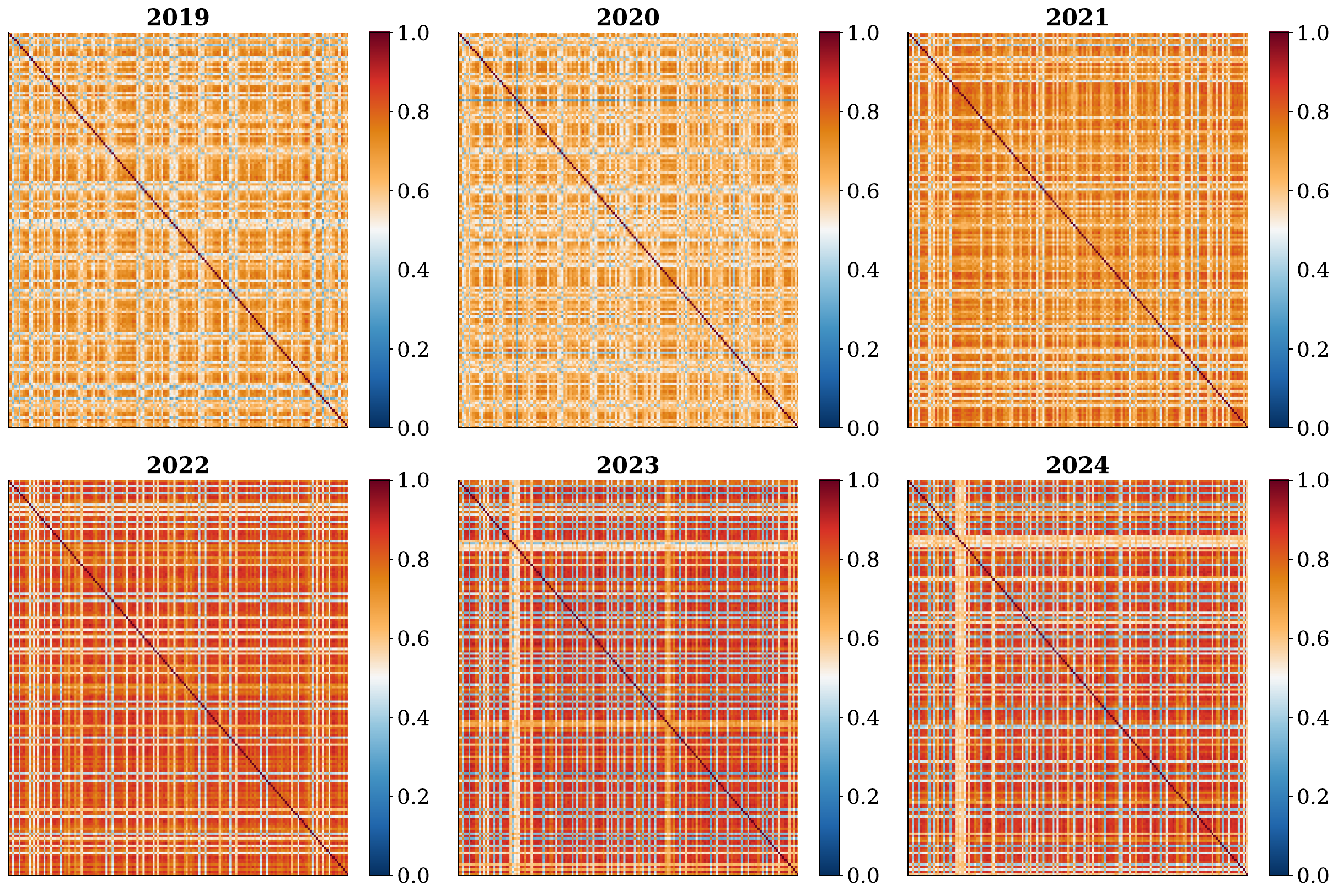}
    \caption{Illustration of the posterior co-clustering matrices for 2019–2024.}
    \label{fig:coclustering-matrices}
\end{figure}

As shown in \cref{fig:bubble-chart}, the cluster structure consists of two dominant and persistent clusters across almost the entire observation period 2019–2024. Cluster 2 is a transient cluster that appears exclusively in 2020 and comprises only four observations. The posterior co-clustering matrices confirm that the dominant two-cluster block structure is recovered consistently across MCMC samples and is not an artifact of the point estimate (see \cref{fig:coclustering-matrices}). Notably, the block structure is already sharp and stable in 2019, with high co-clustering probabilities (yellow to red range) within blocks and near-zero probabilities (blue range) between blocks, and this sharpness is maintained throughout the entire observation period. 

Region-level traces (see \cref{fig:region-traces} in \cref{appx:figures} in the appendix) over time, averaged across topics, show that all eleven regions remain clustered in a narrow band between a mean of approximately $0.237$ and $0.283$ throughout the observation period. Anglo America consistently occupies the upper end of this range (peaking in 2024), while the regions occupying the lower end are not consistent over time. Reports about the Russian and the Oriental region are placed at the lower end in 2019-2021. In 2023-2024, the lower end is composed of reports on South Asia and East Asia. Despite this ordering, all regions exhibit a common upward drift from 2020 to 2022.

However, the impression of an improving (albeit slow) media landscape in terms of gender parity in the sources is deceptive. We define regimes as groups of temporal trend structures in observations (see the definitions and characteristics in \cref{tab:regime-caracteristics-year}).
Regime classification at the label level corroborates the stability finding at the aggregate level. Of the 165 label-time series examined, $86.1\%$ fall into the 'persistent low' regime, characterized by low transition rates and no discernible linear trend. A further $7.9\%$ are classified as 'improving',  while $6.1\%$ show 'volatile' cluster assignments across years. Notably, no worsening observations are found at the region level. As listed in \cref{tab:regime-caracteristics-improving}, the improving labels are concentrated in topics with an overall higher mean cluster membership (topic 2 - public health/healthcare, topic 5 - culture, topic 10 - pandemic, topic 12 - development, topic 14 - travel), and their improvement is predominantly characterized by a transition from cluster 1 to cluster 0 in the later years (2022-2024). Furthermore, the means of the gender bias measurements for the single observation in the 'improving' cluster is higher than the mean of the dominant cluster 1 ($=0.25$). Taken together, we observe a gradual upward drift in a subset of moderate gender-biased region-topic combinations rather than broad-based improvement.

\begin{table}[]
\centering
\caption{Regime Characteristics. OLS: Ordinary Least Square} \label{tab:regime-caracteristics-year}
\begin{tabular}{p{7.5cm}llp{3cm}}
\hline
\textbf{characteristics}                                    & \textbf{count}& \textbf{pct}  & \textbf{label} \\ \hline
transition  rate $>0.5$                                     & 10            & 6.1 \%        & 'Volatile'                    \\
OLS slope $>0.01$ and transition rate $\leq 0.5$            & 13            & 7.9 \%        & 'Improving'                   \\ 
Stable low proportion of female quotes; no temporal trend   & 142           & 86.1 \%       & 'Persistent low'              \\ \hline
\end{tabular}
\end{table}

\begin{table}[]
\centering
\caption{'Improving' regime content with cluster allocation for each time point.} \label{tab:regime-caracteristics-improving}
\begin{tabular}{lp{0.75cm}p{0.75cm}p{0.75cm}p{0.75cm}p{0.75cm}p{0.75cm}ll}
\hline
\textbf{region $\times$ topic}          & \textbf{2019} & \textbf{2020} & \textbf{2021} & \textbf{2022} & \textbf{2023} & \textbf{2024} & \textbf{mean} & \textbf{slope}          \\ \hline
Russian\_T13            & 1             & 1             & 1             & 1             & 1             & 0             & 0.258        & 0.017                               \\
AustralianOceanian\_T11 & 1             & 1             & 1             & 1             & 0             & 0             & 0.271        & 0.016                                  \\
AustralianOceanian\_T14 & 1             & 1             & 1             & 1             & 0             & 0             & 0.275        & 0.012                                  \\
Oriental\_T14           & 1             & 1             & 1             & 1             & 0             & 0             & 0.277        & 0.014                                 \\
EasternEurope\_T10      & 1             & 1             & 1             & 1             & 0             & 0             & 0.277        & 0.016                              \\
Oriental\_T10           & 1             & 1             & 1             & 1             & 0             & 0             & 0.282        & 0.015                               \\
SubSaharanAfrican\_T14  & 1             & 1             & 1             & 0             & 0             & 1             & 0.285        & 0.011                                  \\
EasternEurope\_T12      & 1             & 1             & 1             & 0             & 0             & 0             & 0.289        & 0.020                           \\
LatinAmerican\_T12      & 1             & 1             & 1             & 0             & 0             & 0             & 0.293        & 0.014                            \\
EasternEurope\_T2       & 0             & 1             & 1             & 0             & 0             & 0             & 0.295        & 0.014               \\
AngloAmerican\_T12      & 1             & 1             & 1             & 0             & 0             & 0             & 0.296        & 0.013                           \\
Russian\_T5             & 1             & 1             & 1             & 0             & 0             & 0             & 0.297        & 0.018                                \\
EasternEurope\_T5       & 0             & 0             & 1             & 0             & 0             & 0             & 0.311        & 0.010                                \\ \hline
\end{tabular}
\end{table}

\subsection{Dynamic Density Estimation}\label{sec:density_estimation}
\begin{figure}[ht]
    \centering
    \includegraphics[width=\linewidth]{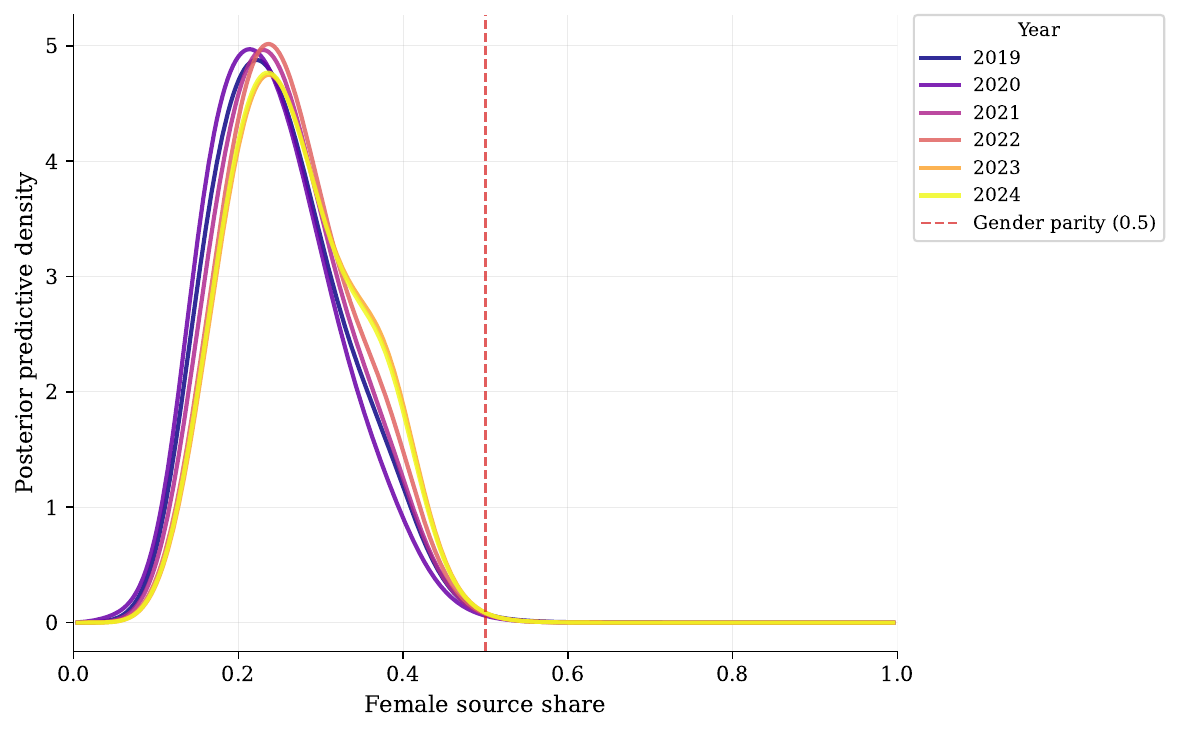}
    \caption{Plot of the density estimation aggregated to the single years.}\label{fig:aggregate-density}
\end{figure}

Beyond the cluster level summaries, the year-level aggregate posterior predictive density of our gender bias measurement stated in equation \eqref{eqn:gender-bias-measurement-observation} (see \cref{fig:aggregate-density}) exhibits two key features of the model's distributional fit. First, the marginal predictive distribution is strongly right-skewed, with a sharp mode at approximately $0.24$-$0.26$ and a long right tail that decays to near zero well before the gender parity threshold of $0.5$. This finding confirms the pervasive, structurally stable under-representation of female sources documented in the clustering analysis in \cref{sec:cluster-structure}. Second, neither the mode, the spread, nor the shape of the predictive distribution shifts meaningfully between 2019 and 2024. The temporal overlap constitutes evidence that the aggregate gender bias in Canadian news has not changed over the observation period. 


\section{Discussion}\label{sec:discussion}
\textbf{Key Findings.}
Our analysis of the cluster structure (see \cref{sec:cluster-structure}) in the Canadian news reporting shows that the proportion of women as sources consistently remains below gender parity, averaging 0.19-0.31 across all regions, topics, and years. The \textit{structural under-representation of female voices in the Canadian news} remains stable throughout the entire observation period from 2019 to 2024. The finding is consistent with existing gender bias studies (e.g., \cite{hooghe2015enduring, jia2016women, rao2021gender, riedl2024journalistic, shor2015paper, shor2019large, taboada2025reported, vanderpas2020gender}), which all show the under-representation of women in the news. The cluster analysis identifies three clusters: cluster 2, which exists only in 2020; cluster 1, which represents the dominant group; and cluster 0, which has the highest proportion of women and primarily encompasses topics such as public health/healthcare and culture. Our results are also consistent with those of \citet{rao2021gender}. The authors identify lifestyle, healthcare, and medical research as female-dominated topics in terms of quote share (based on the same data; see \cref{sec:data}). However, our results reveal a significant difference: despite a higher share of female quotes than in other topics, these topics still remain male-dominated at our level of aggregation. However, it can be concluded that the \textit{cluster structure is topic-driven} and that the influence of the cultural regions being reported on is only recognizable to a limited extent.   

\begin{figure}[H]
    \centering
    \includegraphics[width=\linewidth]{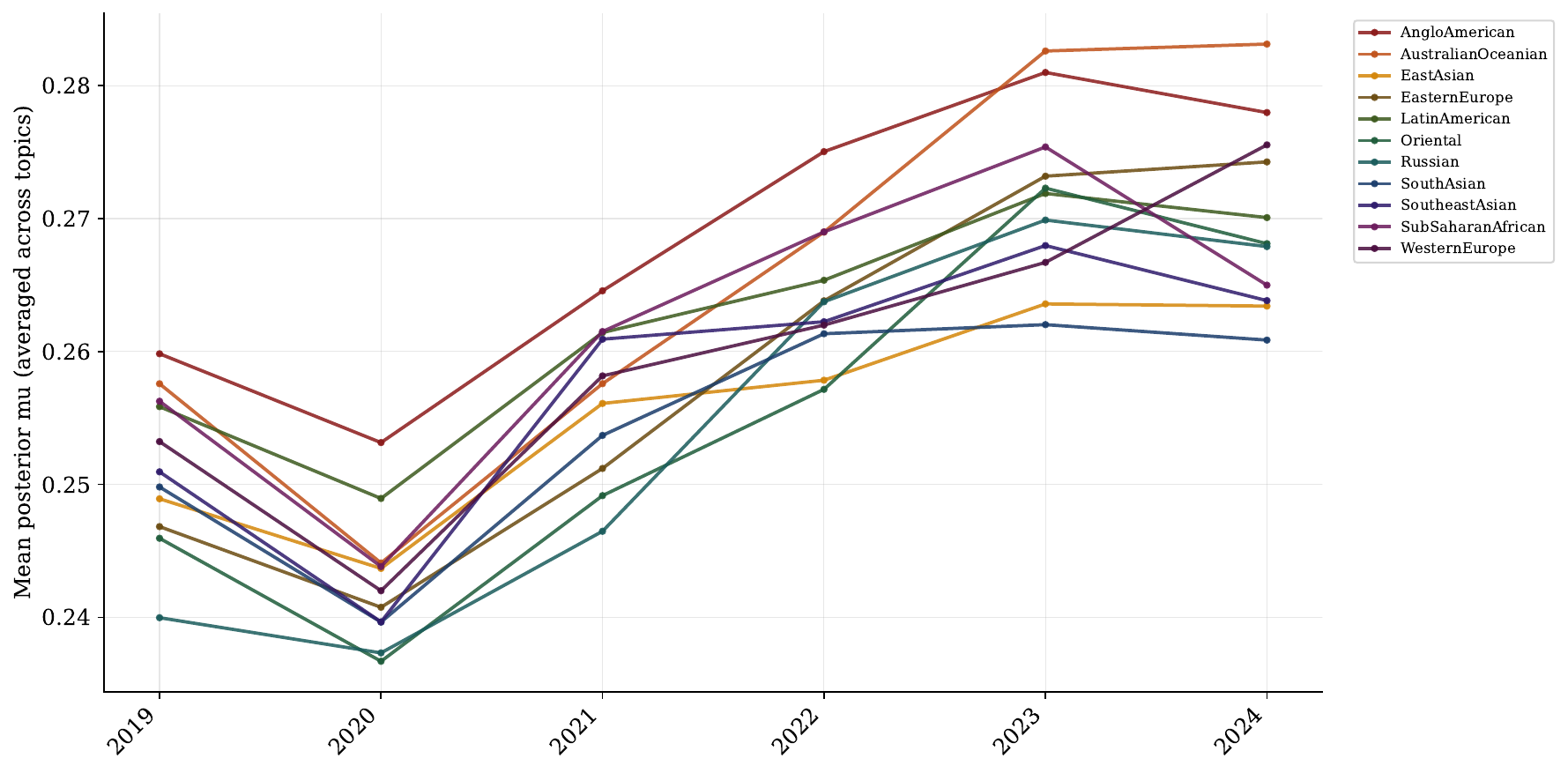}
    \caption{Region-level traces over time. We averaged the posterior mean across topics.}
    \label{fig:region-traces}
\end{figure}

The temporal analysis (see \cref{sec:temporal_dynamics}) shows that only a specific subset of region-topic combinations exhibits a moderate improvement, characterized by observations that are already relatively close to gender parity in the latent space compared to other observations. For instance, reports about Eastern Europe or Russia about culture, or reports about Anglo America about development (see \cref{tab:regime-caracteristics-improving}). The majority of observations fall within the ‘persistent low’ regime with no notable time trend. Furthermore, the density estimate (see \cref{sec:density_estimation}) confirms that the aggregate gender bias remains stable across time, with no significant changes in the distribution of the proportion of women. These results point to a \textit{slow, long-lasting change that is largely limited to specific topics and regions}, without achieving broader improvement in gender parity in Canadian news. This strong stability over time is in line with similar studies reporting a very slow trend toward gender parity (e.g., \cite{deiorio2023bayesian, joshi2020violators, rao2021gender}). A longer study period (e.g., as in \citet{deiorio2023bayesian}, which examines media trends over 100 years) would reveal long-term trends.  Nevertheless, the emergence of cluster 2 in 2020 (see \cref{fig:bubble-chart}), the dip in the region-level traces in 2020 (see \cref{fig:region-traces}), and the beginning shift in color after 2020 in the co-clustering matrices (see \cref{fig:coclustering-matrices} and \cref{fig:coclustering-matrix-country}) may reflect \textit{COVID-19-related disruptions in coverage patterns}.   
\\
\\
\textbf{General Discussion.}
As discussed in \cref{sec:introduction}, gender bias in media coverage reflects societal structures. \citet{rao2021gender} have already noted in their work that it is very difficult to quantitatively assess the distribution of, for example, female experts in specific subject areas or regions. However, since gender bias conversely also influences social structures \cite{hooghe2015enduring, riedl2024journalistic, shor2015paper, vanderpas2020gender}, gender parity in reporting can lead to greater equality, e.g., by increasing the number of women in male-dominated fields. Therefore, we contribute to research on the under-representation of women in Western media (represented by Canada) by highlighting the existing gaps, even though our project does not quantify the reasons for gender bias. 

From a methodological perspective, we used a DP mixture approach with a Beta distribution as the mixture kernel, which, to the best of our knowledge, is the first such application in a dynamic setting. The choice for a Beta distribution respects the bounded support of our target variable stated in equation \eqref{eqn:gender-bias-measurement-observation}. Nevertheless, this decision comes with higher computational costs, since estimating the Beta distribution's parameter requires sampling using MH. 

We summarize that we confirmed existing findings in gender bias research by making the data's underlying probabilistic structure explicit and analytically accessible through our adapted AR1-DP model, thereby avoiding the need for strong assumptions about the underlying data structure. 
Our project bridges the gap between social science research on gender bias and an advanced statistical approach capable of adequately capturing and addressing its complexity. This balancing act leads to one of the main limitations of our work. We conduct our analysis at a very broad level of aggregation and examine gender bias in Canadian news solely in terms of the coverage of female voices in articles. Thereby, we neglect the perspective on the content at the text level. For instance, we do not analyze the specific context in which women are quoted (e.g., major global sport events vs. small local events), nor do we examine the level of language (e.g., the framing).
Nevertheless, our work provides an important early-stage validation of the use of more advanced probabilistic models for analyzing bias in media. Much existing research on media bias relies on descriptive content analysis or focuses on tasks such as media bias classification \cite{horych2025promises, spinde2025automated}. Our results suggest that more complex statistical frameworks can reproduce established findings while additionally capturing latent dependencies and structural relations within media bias data. We therefore see our study not only as a substantive contribution to gender bias research, but also as a methodological proof of concept that may encourage future work to investigate different forms of bias in the media using more advanced probabilistic and model-based approaches.
For example, there is less research on the use of spatio-temporal statistical models for answering media bias-related research questions \cite{habereder2025systematic}. Therefore, future research in applied spatio-temporal statistics should consider different forms of media bias to uncover the underlying structures of one of the most relevant issues for society.


\section{Conclusion}\label{sec:conclusion}
To address the complexity of gender bias, we aim to develop advanced statistical methods. Our project is motivated by the work of \citet{deiorio2023bayesian} on recent methodological developments in DP models for time-dependent gender-stereotype data, as well as by the work of \citet{rao2021gender}, which contributes to gender bias research by uncovering gender inequality in media representation. In our project, we measure gender bias in terms of coverage by the proportion of females quoted in an article. We grouped and aggregated our data, a set of Canadian news articles, by narrative location (i.e., the geographical region the article reports on), article topic, and year of publication. We used a time-dependent Bayesian nonparametric Dirichlet process mixture model with a Beta distribution as the mixture kernel to perform dynamic clustering and density estimation. We find a topic-driven cluster structure and slow, long-lasting temporal dynamics that are largely limited to specific topics and narrative locations. Moreover, we can confirm the structural under-representation of women in the media, a phenomenon well documented in the literature, including in the coverage of female voices. Balanced gender representation in the news also influences gender equality within society. Achieving a society in which the genders are equal will require a great deal of effort and patience, because, as our findings show, change is happening slowly.      

\section*{Funding}
This work was supported by a fellowship of the German Academic Exchange Service (DAAD) and the Hanns-Seidel Foundation. This work was partially supported by JSPS KAKENHI Grants JP21H04907 and JP24H00732, by JST CREST Grants JPMJCR20D3 and JPMJCR2562 including AIP challenge program, by JST AIP Acceleration Grant JPMJCR24U3, and by JST K Program Grant JPMJKP24C2 Japan, and by the Deutsche Forschungsgemeinschaft (DFG, German Research Foundation) under project number 565115197 (BARI). 

\section*{Disclosure Statement}
The authors report there are no competing interests to declare.

\section*{Data Availability Statement}
The data that support the findings of this study are available from Simon Fraser University (SFU). These data are available under license and subject to restrictions. Researchers interested in accessing the Gender Gap Tracker dataset may contact Prof. Maite Taboada in the Department of Linguistics at SFU.

\section*{Ethics Statement - Use of Generative AI}
We would like to point out that the authors used an AI-based tool (GPT-5.5, OpenAI) to assist in preparing this work, i.e., to revise text passages for minor rephrasing and grammar correction, and to optimize individual code sections. Originality and accuracy of the content have been verified by the authors. The authors have reviewed the applicable terms of use of ChatGPT and confirm that its use is consistent with the publication requirements of this work. The authors take full responsibility for the integrity, originality, and accuracy of the entire manuscript, including all references and citations.

\bibliographystyle{plainnat} 
\bibliography{references}

@article{andrieu2010particle,
  title={Particle Markov Chain Monte Carlo Methods},
  author={Andrieu, Christophe and Doucet, Arnaud and Holenstein, Roman},
  journal={Journal of the Royal Statistical Society Series B: Statistical Methodology},
  volume={72},
  number={3},
  pages={269--342},
  year={2010},
  publisher={Oxford University Press}
}

@article{binder1978bayesian,
  title={Bayesian cluster analysis},
  author={Binder, David A},
  journal={Biometrika},
  volume={65},
  number={1},
  pages={31--38},
  year={1978},
  publisher={Oxford University Press}
}

@article{blei2003latent,
  title={Latent dirichlet allocation},
  author={Blei, David M and Ng, Andrew Y and Jordan, Michael I},
  journal={Journal of machine Learning research},
  volume={3},
  number={Jan},
  pages={993--1022},
  year={2003}
}

@article{castro2021measuring,
  title={Measuring partisan media bias cross-nationally},
  author={Castro, Laia},
  journal={Swiss Political Science Review},
  volume={27},
  number={2},
  pages={412--433},
  year={2021},
  publisher={Wiley Online Library}
}

@article{costa_jussa2019gender,
  author  = {Marta R. Costa-juss{\`a}},
  title   = {An Analysis of Gender Bias Studies in Natural Language Processing},
  journal = {Nature Machine Intelligence},
  volume  = {1},
  pages   = {495--496},
  year    = {2019}
}

@inproceedings{dacon2021gender,
  title={Does gender matter in the news? detecting and examining gender bias in news articles},
  author={Dacon, Jamell and Liu, Haochen},
  booktitle={Companion Proceedings of the Web Conference 2021},
  pages={385--392},
  year={2021}
}

@article{deiorio2023bayesian,
  title={Bayesian nonparametric mixture modeling for temporal dynamics of gender stereotypes},
  author={De Iorio, Maria and Favaro, Stefano and Guglielmi, Alessandra and Ye, Lifeng},
  journal={The Annals of Applied Statistics},
  volume={17},
  number={3},
  pages={2256--2278},
  year={2023},
  publisher={Institute of Mathematical Statistics}
}

@article{elmasry2025gender,
  title={Gender hierarchies in reporting genocide: an analysis of the dehumanization of Palestinian men in Western media},
  author={El Masry, Noura and Sawaf, Zina and King, Gretchen and Baroudi, Sami},
  journal={Communication, Culture \& Critique},
  volume={18},
  number={4},
  pages={310--321},
  year={2025},
  publisher={Oxford University Press}
}

@article{ferguson1973bayesian,
  title={A bayesian analysis of some non-parametric problems},
  author={Ferguson, Thomas S},
  journal={The Annals of Statistics},
  volume={1},
  number={2},
  pages={353--355},
  year={1873}
}

@article{greenwald2023israeli,
  title={Israeli media coverage of international male and female politicians: Gender and ethnopolitical aspects},
  author={Greenwald, Gilad},
  journal={Communications},
  volume={48},
  number={2},
  pages={226--248},
  year={2023},
  publisher={De Gruyter}
}

@article{griffin2011stick,
  title={Stick-breaking autoregressive processes},
  author={Griffin, Jim E and Steel, Mark FJ},
  journal={Journal of econometrics},
  volume={162},
  number={2},
  pages={383--396},
  year={2011},
  publisher={Elsevier}
}

@article{griffin2006order,
  title={Order-based dependent Dirichlet processes},
  author={Griffin, Jim E and Steel, MF J},
  journal={Journal of the American statistical Association},
  volume={101},
  number={473},
  pages={179--194},
  year={2006},
  publisher={Taylor \& Francis}
}

@article{gutierrez2016airquality,
  title={A time dependent Bayesian nonparametric model for air quality analysis},
  author={Guti{\'e}rrez, Luis and Mena, Rams{\'e}s H and Ruggiero, Matteo},
  journal={Computational Statistics \& Data Analysis},
  volume={95},
  pages={161--175},
  year={2016},
  publisher={Elsevier}
}

@article{habereder2025systematic,
  title={A Systematic Review of Spatio-Temporal Statistical Models: Theory, Structure, and Applications},
  author={Habereder, Isabella and Kneib, Thomas and Echizen, Isao and Spinde, Timo},
  journal={arXiv preprint arXiv:2511.00422},
  year={2025}
}

@article{hooghe2015enduring,
  title={Enduring gender bias in reporting on political elite positions: Media coverage of female MPs in Belgian news broadcasts (2003--2011)},
  author={Hooghe, Marc and Jacobs, Laura and Claes, Ellen},
  journal={The International Journal of Press/Politics},
  volume={20},
  number={4},
  pages={395--414},
  year={2015},
  publisher={SAGE Publications Sage CA: Los Angeles, CA}
}

@inproceedings{horych2025promises,
  title={The promises and pitfalls of LLM annotations in dataset labeling: A case study on media bias detection},
  author={Horych, Tom{\'a}{\v{s}} and Mandl, Christoph and Ruas, Terry and Greiner-Petter, Andr{\'e} and Gipp, Bela and Aizawa, Akiko and Spinde, Timo},
  booktitle={Findings of the Association for Computational Linguistics: NAACL 2025},
  pages={1370--1386},
  year={2025}
}

@article{jia2016women,
  title={Women are seen more than heard in online newspapers},
  author={Jia, Sen and Lansdall-Welfare, Thomas and Sudhahar, Saatviga and Carter, Cynthia and Cristianini, Nello},
  journal={PloS one},
  volume={11},
  number={2},
  pages={e0148434},
  year={2016},
  publisher={Public Library of Science San Francisco, CA USA}
}

@article{joshi2020violators,
  title={Violators, virtuous, or victims? How global newspapers represent the female member of parliament},
  author={Joshi, Devin K and Hailu, Meseret F and Reising, Lauren J},
  journal={Feminist Media Studies},
  volume={20},
  number={5},
  pages={692--712},
  year={2020},
  publisher={Taylor \& Francis}
}

@inproceedings{joulin2017bag,
  title={Bag of tricks for efficient text classification},
  author={Joulin, Armand and Grave, Edouard and Bojanowski, Piotr and Mikolov, Tom{\'a}{\v{s}}},
  booktitle={Proceedings of the 15th conference of the European chapter of the association for computational linguistics: volume 2, short papers},
  pages={427--431},
  year={2017}
}

@article{joulin2016fasttext,
  title={Fasttext. zip: Compressing text classification models},
  author={Joulin, Armand and Grave, Edouard and Bojanowski, Piotr and Douze, Matthijs and J{\'e}gou, H{\'e}rve and Mikolov, Tomas},
  journal={arXiv preprint arXiv:1612.03651},
  year={2016}
}

@incollection{kolb1962,
  author    = {Albert Kolb},
  title     = {Die Geographie und die Kulturerdteile},
  booktitle = {Hermann von Wissmann-Festschrift},
  editor    = {A. Leidlmair},
  publisher = {Geographisches Institut der Universität Tübingen},
  year      = {1962},
  pages     = {46}
}

@inproceedings{kottas2006dirichlet,
  title={Dirichlet process mixtures of beta distributions, with applications to density and intensity estimation},
  author={Kottas, Athanasios},
  booktitle={Workshop on Learning with Nonparametric Bayesian Methods, 23rd International Conference on Machine Learning (ICML)},
  volume={47},
  year={2006}
}

@article{kottas2007spatial,
  title={Bayesian mixture modeling for spatial Poisson process intensities, with applications to extreme value analysis},
  author={Kottas, Athanasios and Sans{\'o}, Bruno},
  journal={Journal of Statistical Planning and Inference},
  volume={137},
  number={10},
  pages={3151--3163},
  year={2007},
  publisher={Elsevier}
}

@article{lau2007bayesian,
  title={Bayesian model-based clustering procedures},
  author={Lau, John W and Green, Peter J},
  journal={Journal of Computational and Graphical Statistics},
  volume={16},
  number={3},
  pages={526--558},
  year={2007},
  publisher={Taylor \& Francis}
}

@article{lee2025logistic,
  title={Logistic-beta processes for dependent random probabilities with beta marginals},
  author={Lee, Changwoo J and Zito, Alessandro and Sang, Huiyan and Dunson, David B},
  journal={Bayesian Analysis},
  volume={20},
  number={4},
  pages={1345--1369},
  year={2025},
  publisher={International Society for Bayesian Analysis}
}

@article{lee2024visuality,
  title={The specific visuality of women of the global South in the media of the global North},
  author={Lee, Sohyun},
  journal={Humanities and Social Sciences Communications},
  volume={11},
  number={1},
  pages={1--10},
  year={2024},
  publisher={Palgrave}
}

@article{lo1984class,
  title={On a class of Bayesian nonparametric estimates: I. Density estimates},
  author={Lo, Albert Y},
  journal={The annals of statistics},
  pages={351--357},
  year={1984},
  publisher={JSTOR}
}

@article{marin2026bayesian,
  title={Bayesian nonparametric modeling of dynamic pollution clusters through an autoregressive logistic-beta Stirling-gamma process},
  author={Marin, Santiago and Loong, Bronwyn and Westveld, Anton H},
  journal={arXiv preprint arXiv:2601.04625},
  year={2026}
}

@article{meila2007comparing,
  title={Comparing clusterings—an information based distance},
  author={Meil{\u{a}}, Marina},
  journal={Journal of multivariate analysis},
  volume={98},
  number={5},
  pages={873--895},
  year={2007},
  publisher={Elsevier}
}

@misc{newig2014,
  author       = {J. Newig},
  title        = {Das Konzept der Kulturerdteile},
  year         = {2014},
  institution  = {Universität Kiel},
  address      = {Kiel},
  url          = {http://www.kulturerdteile.de/kulturerdteile/},
  note         = {Last access on 24.03.2026}
}

@article{page2022dependent,
  title={Dependent modeling of temporal sequences of random partitions},
  author={Page, Garritt L and Quintana, Fernando A and Dahl, David B},
  journal={Journal of Computational and Graphical Statistics},
  volume={31},
  number={2},
  pages={614--627},
  year={2022},
  publisher={Taylor \& Francis}
}

@article{power2019women,
  author  = {Kate Power and Lucy Rak and Marianne Kim},
  title   = {Women in Business Media: A Critical Discourse Analysis of Representations of Women in Forbes, Fortune and Bloomberg BusinessWeek, 2015--2017},
  journal = {Critical Approaches to Discourse Analysis Across Disciplines},
  volume  = {11},
  number  = {2},
  pages   = {1--26},
  year    = {2019}
}

@article{rao2021gender,
  title={Gender bias in the news: A scalable topic modelling and visualization framework},
  author={Rao, Prashanth and Taboada, Maite},
  journal={Frontiers in artificial intelligence},
  volume={4},
  pages={664737},
  year={2021},
  publisher={Frontiers Media SA}
}

@article{riedl2024journalistic,
  title={“I can’t just pull a woman out of a hat”: A mixed-methods study on journalistic drivers of women’s representation in political news},
  author={Riedl, Andreas A and Rohrbach, Tobias and Krakovsky, Christina},
  journal={Journalism \& Mass Communication Quarterly},
  volume={101},
  number={3},
  pages={679--702},
  year={2024},
  publisher={SAGE Publications Sage CA: Los Angeles, CA}
}

@article{robbins1951stochastic,
  title={A stochastic approximation method},
  author={Robbins, Herbert and Monro, Sutton},
  journal={The annals of mathematical statistics},
  pages={400--407},
  year={1951},
  publisher={JSTOR}
}

@article{roberts2001optimal,
  title={Optimal scaling for various Metropolis-Hastings algorithms},
  author={Roberts, Gareth O and Rosenthal, Jeffrey S},
  journal={Statistical science},
  volume={16},
  number={4},
  pages={351--367},
  year={2001},
  publisher={Institute of Mathematical Statistics}
}

@article{sethuraman1994constructive,
  title={A constructive definition of Dirichlet priors},
  author={Sethuraman, Jayaram},
  journal={Statistica sinica},
  pages={639--650},
  year={1994},
  publisher={JSTOR}
}

@article{shor2019women,
  title={Do women in the newsroom make a difference? Coverage sentiment toward women and men as a function of newsroom composition},
  author={Shor, Eran and Van de Rijt, Arnout and Miltsov, Alex},
  journal={Sex Roles},
  volume={81},
  number={1},
  pages={44--58},
  year={2019},
  publisher={Springer}
}

@article{shor2015paper,
  title={A paper ceiling: Explaining the persistent underrepresentation of women in printed news},
  author={Shor, Eran and Van De Rijt, Arnout and Miltsov, Alex and Kulkarni, Vivek and Skiena, Steven},
  journal={American Sociological Review},
  volume={80},
  number={5},
  pages={960--984},
  year={2015},
  publisher={Sage publications Sage CA: Los Angeles, CA}
}

@article{shor2019large,
  title={A large-scale test of gender bias in the media},
  author={Shor, Eran and Van De Rijt, Arnout and Fotouhi, Babak},
  journal={Sociological science},
  volume={6},
  pages={526--550},
  year={2019}
}

@article{smithson2006better,
  title={A better lemon squeezer? Maximum-likelihood regression with beta-distributed dependent variables.},
  author={Smithson, Michael and Verkuilen, Jay},
  journal={Psychological methods},
  volume={11},
  number={1},
  pages={54},
  year={2006},
  publisher={American Psychological Association}
}

@book{spinde2025automated,
  title={Automated Detection of Media Bias: From the Conceptualization of Media Bias to its Computational Classification},
  author={Spinde, Timo},
  year={2025},
  publisher={Springer Nature}
}

@article{taboada2025reported,
  title={Reported speech and gender in the news: Who is quoted, how are they quoted, and why it matters},
  author={Taboada, Maite},
  journal={Discourse \& Communication},
  volume={19},
  number={1},
  pages={93--113},
  year={2025},
  publisher={SAGE Publications Sage UK: London, England}
}

@article{trimble2021gender,
  title={Gender novelty and personalized news coverage in Australia and Canada},
  author={Trimble, Linda and Curtin, Jennifer and Wagner, Angelia and Auer, Meagan and Woodman, VKG and Owens, Bethan},
  journal={International Political Science Review},
  volume={42},
  number={2},
  pages={164--178},
  year={2021},
  publisher={SAGE Publications Sage UK: London, England}
}

@article{vanderpas2020gender,
  title={Gender differences in political media coverage: A meta-analysis},
  author={Van der Pas, Daphne Joanna and Aaldering, Loes},
  journal={Journal of Communication},
  volume={70},
  number={1},
  pages={114--143},
  year={2020},
  publisher={Oxford University Press}
}

@article{wade2018bayesian,
  title={Bayesian Cluster Analysis: Point Estimation and Credible Balls (with Discussion)},
  author={Wade, Sara and Ghahramani, Zoubin},
  journal={Bayesian Analysis},
  volume={13},
  number={2},
  pages={559--626},
  year={2018}
}

\newpage
\section{Appendix}\label{appendix}
In \cref{appx:figures} we present the figures mentioned in \cref{sec:application}. Furthermore, we use the appendix to describe the data pre-processing in \cref{appx:data-preprocess}. In \cref{appx:paramtuning}, we present the parameter selection and validation of the final model used to obtain the results in \cref{sec:application}. In \cref{appx:country-level} we describe the results obtained for the country $\times$ topic $\times$ year aggregation level. Our code for the model is publicly available in \url{https://github.com/Media-Bias-Group/Nonparametric-Bayesian-Mixture-Model-for-Gender-Bias-Analysis}. 

\subsection{Figures}\label{appx:figures}
\begin{figure}[H]
    \centering
    \includegraphics[width=\linewidth]{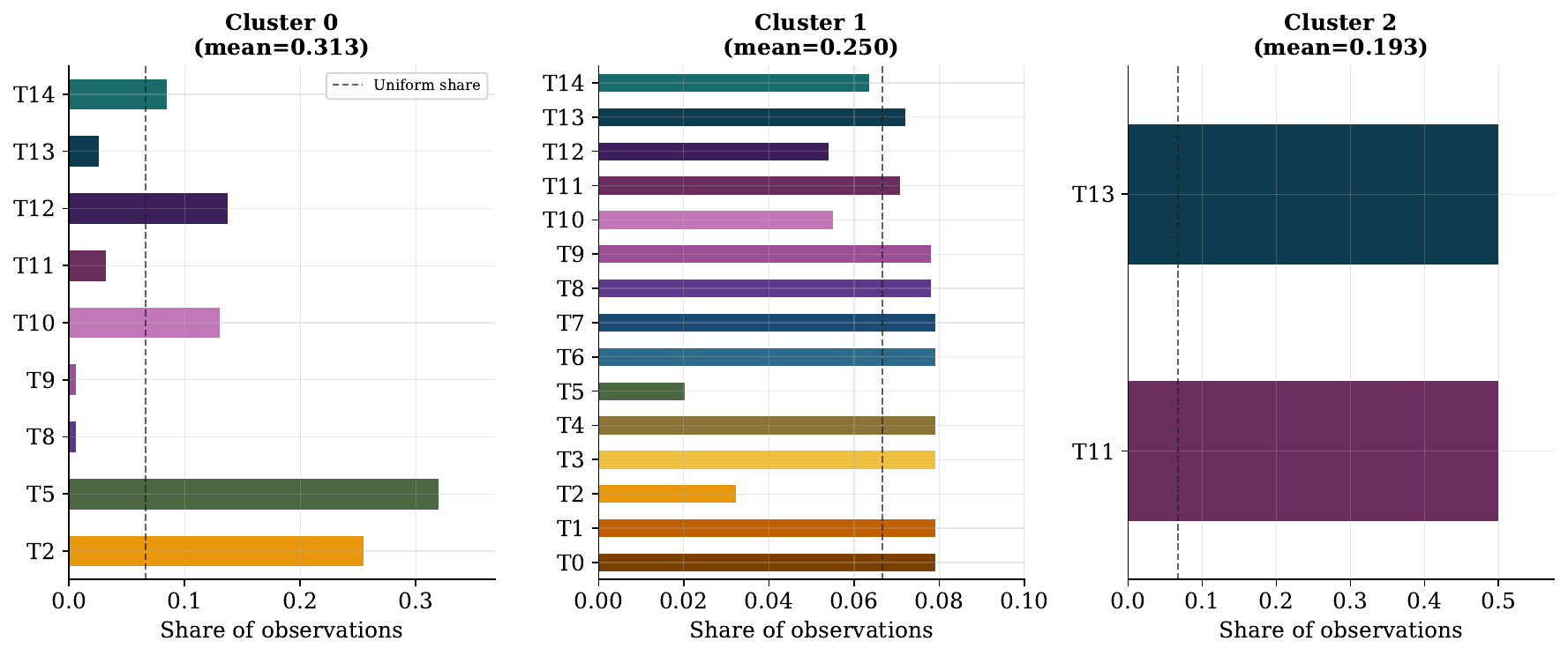}
    \caption{Topic composition per cluster}
    \label{fig:topic-composition}
\end{figure}

\begin{figure}[H]
    \centering
    \includegraphics[width=\linewidth]{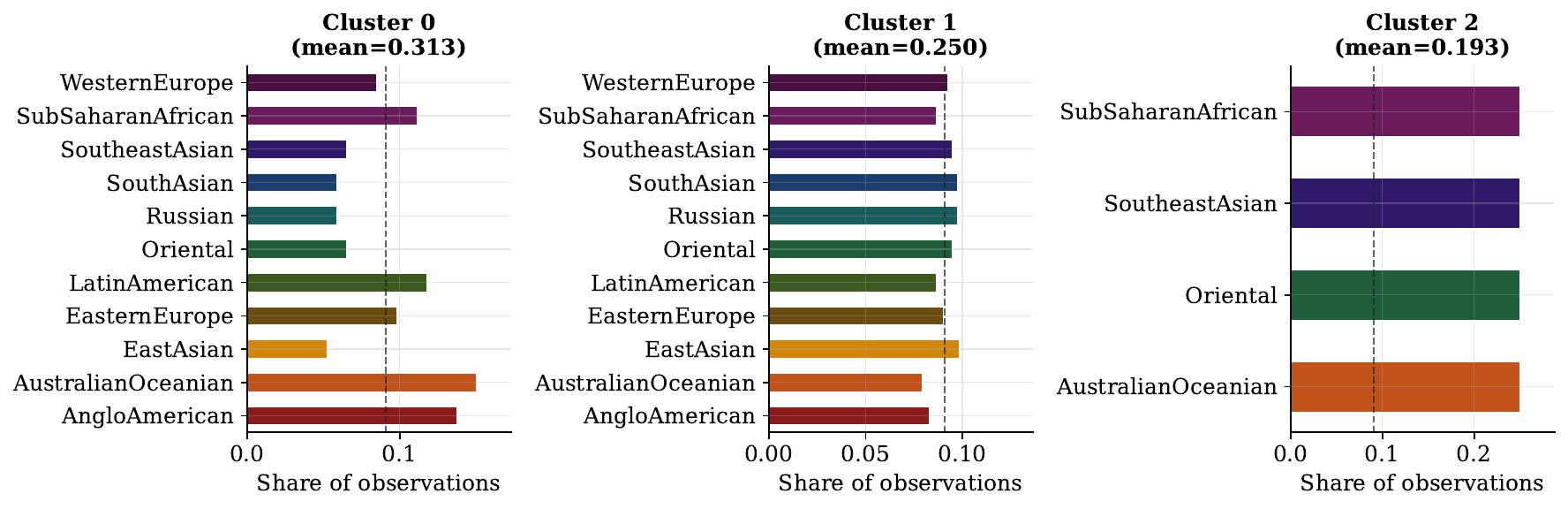}
    \caption{Region composition per cluster}
    \label{fig:region-composition}
\end{figure}

\begin{figure}[H]
    \centering
    \includegraphics[width=\linewidth]{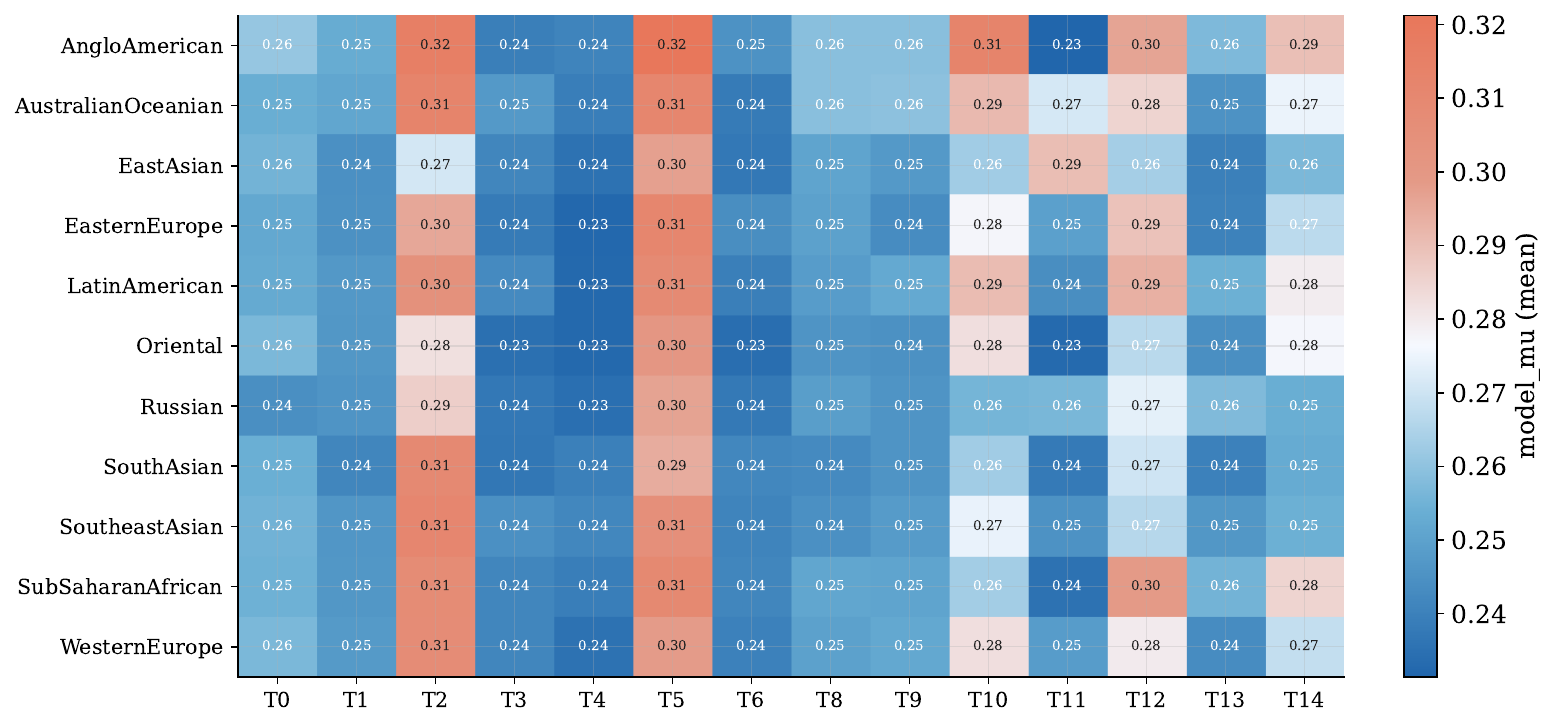}
    \caption{Region $\times$ topic heatmap of posterior mean across all time points.}
    \label{fig:region-topic-heatmap}
\end{figure}

\begin{figure}[H]
    \centering
    \includegraphics[width=\linewidth]{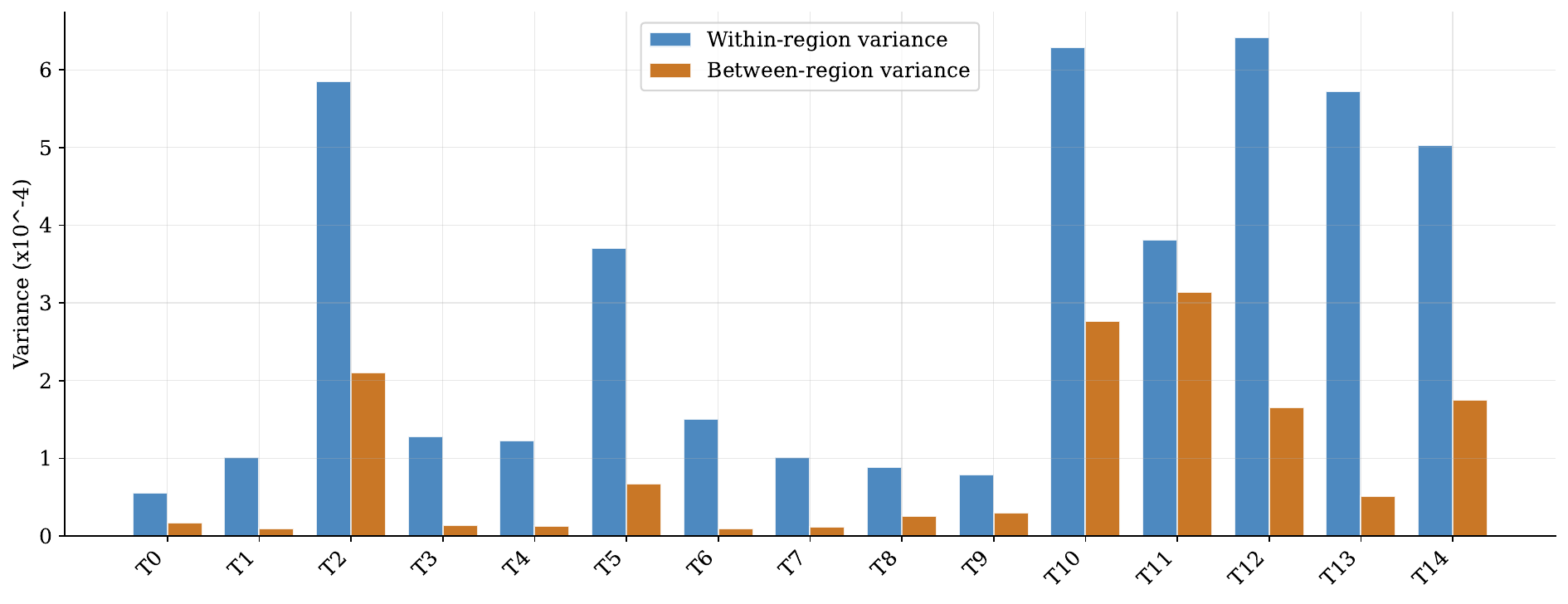}
    \caption{Illustration of the between- vs. within-region variance by topic.}
    \label{fig:variance-analysis}
\end{figure}

\subsection{Data Pre-Pocessing}\label{appx:data-preprocess}
The dataset consists of English- and French-language text. First, we determined each article’s language by using a pretrained language classifier \textit{fastText} \cite{joulin2016fasttext, joulin2017bag}. We only included English-language news articles in our analysis. 
We extended the narrative location for each article. We define narrative locations (in the following also referred to as \textit{country}) as the geopolitical entities (GPE) referenced within an article. We extracted the narrative locations of the articles by employing the following pipeline. First, we used the \textit{spaCy}\footnote{\url{https://spacy.io/}} named-entity recognition model to identify all article-level entities. We then filtered by their class tag to exclude all non-GPE entities (e.g., persons, events, organizations). Identified countries are then resolved by querying a nominatim database\footnote{\url{https://nominatim.org/}}, which returns a canonical name and centroid latitude and longitude. 
We identified each article’s most important topic by employing Latent Dirichlet Allocation\cite{blei2003latent}. To restrict computational demands, we used the first 500 words of every article as input to our model. We processed each text by tokenizing, removing common stop words, lemmatizing, and then assigning the most probable topic. We tested models with 10–20 topics and assigned interpretable topic labels by inspecting the 20 highest-weighted terms in each topic. In line with \citet{rao2021gender}, we have decided to use 15 topics (see \cref{tab:topics} for a detailed description). 
We performed an initial cleaning step by removing all rows with NAN values in the columns \textit{number of female quotes} and \textit{number of male quotes}. We also removed all rows with invalid date representation.  
We constructed a new variable, GenderBias (equation \eqref{eqn:gender-bias-measurement-article}). This variable serves as our measure of gender bias (see \cref{sec:introduction}). 
Due to insufficient data coverage for 2018 and 2025, we limited the analysis to the full years 2019 to 2024.
Next, we performed a column-cleaning step, removing all columns containing metadata we do not need for our analysis. Since we only examine Canadian news articles, we assume that all articles that are not assigned a narrative location report on Canada itself. Therefore, we artificially assigned the narrative location of Canada to each such article. 
We performed two different aggregation levels. First, we aggregated the data at a country $\times$ topic $\times$ year level. Second, we grouped the countries into cultural regions (hereafter referred to as \textit{regions}), motivated by the work of \citet{kolb1962} and \citet{newig2014}, and aggregated the data at the region $\times$ topic $\times$ year level. For our analysis, we only include the combinations with more than two observations aggregated. All included countries and cultural regions are listed in \cref{tab:list-of-regions-and-countries}. The gender bias measurement is aggregated regarding \eqref{eqn:gender-bias-measurement-observation}. Since the distributions used in our models (see \cref{sec:methodology}) operate on the open interval, we transformed the entries of the target variable to (0,1) \cite{smithson2006better}.
To ensure a sufficiently balanced panel structure and the statistical validity of the analysis, we filtered the data based on time-series completeness. Excluding highly incomplete or sporadic time series reduces distortions caused by missing data, increases comparability between units, and ensures stable parameter estimates. Therefore, we only kept rows that met the following thresholds: 
\begin{enumerate}
    \item[(A)] The country/region is observed in at least $80\%$ of all years.
    \item[(B)] The country/region exhibits in at least $80\%$ among the observed years a topic coverage of at least $80\%$.
\end{enumerate}


\subsection{Parameter Tuning and Model Validation}\label{appx:paramtuning}
For the model of \cref{sec:methodology}, we kept the parameters $M$ and $\psi$ fixed. The MCMC scheme for sampling from the corresponding posterior distribution is also described in \cref{sec:methodology}. In addition to this simpler model we also tested a model, where the parameters $M$ and $\psi$ are sampled, therefore the goal is to sample from the joint posterior distribution $p(\theta, s_{1:T}, w_{1:T}, \epsilon_{1:T}, \psi, M \vert y_{1:T})$. The design of the Gibbs sampler is an extension of the algorithm described in \cref{sec:methodology} and follows \citet{deiorio2023bayesian}. We sample from $p(\psi, w_{1:T}, \epsilon_{1:T}\vert s_{1:T}, M)$ and $p(M \vert \psi, s_{1:T})$ by using Particle Marginal Metropolis Hastings (PMMH) with Conditional Sequential Monte Carlo (CSMC) \cite{andrieu2010particle}.

For all parameters marked with (+ SA) in \cref{tab:params}, we performed a WAIC sensitivity analysis and selected the parameter that yielded the lowest WAIC. Since the choice of parameters $\psi$ and $M$ is robust and the computational costs for the model with variables $\psi$ and $M$ sampled with PMMH and CSMC are disproportionate to the significance of the result, we have decided to use the ‘simpler’ model for our project. For the truncation level $J$, we performed a simple truncation check by considering the latest weights $w_{t,J}$ for $t=1,\dots,T$ and calculating standard statistics (mean, maximum, $99\%$ percentile). We compare the maximum to the threshold value 0.01. 

We have three types of parameters: sampler parameters, prior hyperparameters, and model structure parameters. We summarize the parameter choices and the tuning methods in \cref{tab:params}.

\begin{table}[htbp]
\centering
\begin{footnotesize}
\caption{Choice of parameters for our analysis. SA: WAIC sensitivity analysis} \label{tab:params}
\begin{tabular}{llccl}
\hline
\multicolumn{1}{c}{\textbf{Parameter}}  &                           &\multicolumn{2}{c}{\textbf{Value}}         & \multicolumn{1}{c}{\textbf{Tuning-Method}}\\ 
                                        &                           & region level  & country level             &           \\ \hline
\multicolumn{4}{l}{\textbf{A: Sampler-Parameter}}                                                                           \\
Iterations                              &                           & \multicolumn{2}{c}{$50\,000$}             & manual    \\
Burn-in                                 &                           & \multicolumn{2}{c}{$25\,000$}             & manual    \\
Thinning-Factor                         &                           & \multicolumn{2}{c}{100}                   & manual    \\
$\theta$-MH SDs                         & $\sigma_\phi$, $\sigma_m$ & \multicolumn{2}{c}{—}                     & adaptive  \\
$\epsilon$-MH SD                        & $\hat{\sigma}_\epsilon$   & \multicolumn{2}{c}{—}                     & adaptive  \\
Robbins-Monro stepsize                  &                           & \multicolumn{2}{c}{C=1.0, $\alpha$=0.6}   & manual    \\ \hline
\multicolumn{4}{l}{\textbf{B: Prior-Hyperparameter}}                                                                        \\
Prior-mean for m          &                           & \multicolumn{2}{c}{Median(logit y)}       & empirical Bayesian                            \\
Prior-SD for m                         &                           & \multicolumn{2}{c}{$1.5 \cdot$SD(logit y)} & empirical Bayesian (+ SA)                     \\
Gamma-Shape for $\phi$                  & $\alpha_0$                & \multicolumn{2}{c}{4.0}                   & manual                                        \\
Gamma-Rate for $\phi$                   &                           & \multicolumn{2}{c}{$\frac{\text{scale}\cdot \alpha_0}{\phi_{\text{MoM}}}$}  & empirical Bayesian     \\
                                        & scale                     & 0.3           &      0.1                     &manual (+ SA)                                  \\ 
                                        &$\phi_{\text{MoM}}$ & \multicolumn{2}{c}{$\frac{\text{Mean(y)}(1-\text{Mean(y)})}{\text{Var(y)}}-1$} & Beta method-of-moments     \\\hline
\multicolumn{4}{l}{\textbf{C: Model structure}}                                                                                     \\ 
J                                       &       & \multicolumn{2}{c}{15}       & truncation check                                         \\
$\psi$                                  &       & 0.85          &  0.75        & manual (+ SA)                                  \\
M                                       &       & 0.4           &  0.5        & manual (+ SA)                                  \\ \hline
\end{tabular}
\end{footnotesize}
\end{table}


\subsection{Application at Country $\times$ Topic $\times$ Year Aggregation: Gender Bias in the Canadian News}\label{appx:country-level}
\citet{wade2018bayesian} showed that cluster estimation using variation of information (VI) offers the advantage over the Binder loss \cite{binder1978bayesian} of penalizing small clusters, thereby leading to a more interpretable estimated partition. Since our analyses have shown that the country $\times$ topic $\times$ year aggregation is more granular than the region $\times$ topic $\times$ year aggregation in \cref{sec:application}, we use to summarize the samples for the cluster allocations $s_{1:T}$ from the posterior distribution, the VI loss \cite{meila2007comparing} with greedy minimization \cite{wade2018bayesian}, evaluated via the posterior co-clustering matrix, to locate the optimal partition.

The country-level analysis both corroborates and extends the cultural region-level findings in \cref{sec:application}. We also find a dominant cluster across countries, confirming that the systematic under-representation of female quotes is not an artifact of geographic aggregation. We also confirm the topic-driven cluster structure. Even if not as pronounced as at the regional level of aggregation. However, the country-level analysis also reveals structural features that are invisible at the regional level. Most notably, two extreme clusters emerge that have no region-level equivalent: one cluster with a mean of  $= 0.001$ and $N=262$ observations included, captures country $\times$ topic combinations with nearly zero female source share, concentrated in topics 11 (Sport), 13 (Mobility), and 4 (Corporate) an on the countries Rwanda, North Korea, Mongolia, Kuwait, Ivory Coast, Greenland, Cyprus, Bosnia and Herzegovina, and Algeria. Another cluster with a mean $= 0.53$ includes $N=23$ observations, identifies a small set of observations that exceed gender parity, a finding entirely obscured by aggregation at the region level. 

\begin{figure}[ht]
    \centering
    \includegraphics[width=\linewidth]{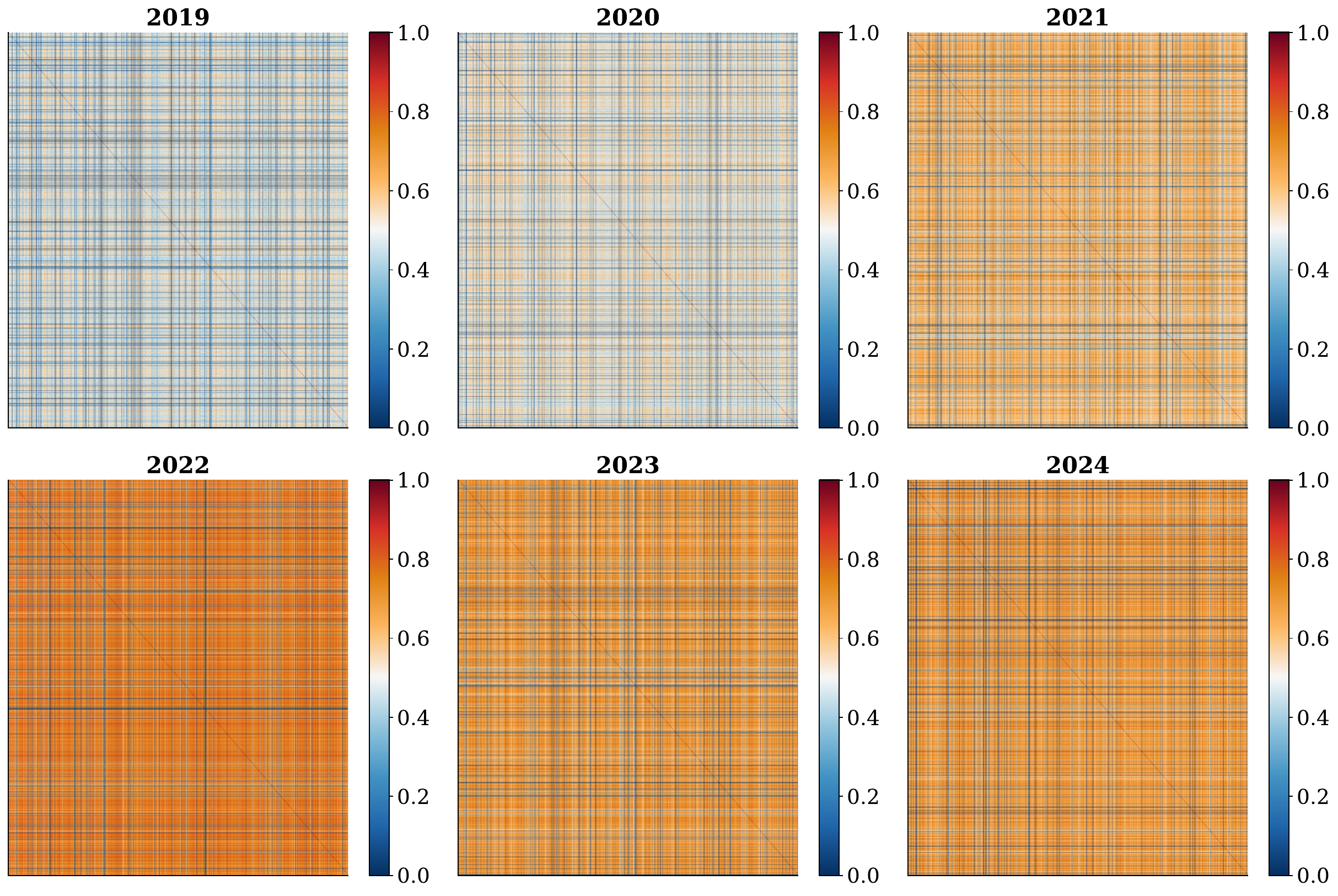}
    \caption{Illustration of the posterior co-clustering matrices for the country $\times$ topic aggregation for 2019-2024}
    \label{fig:coclustering-matrix-country}
\end{figure}

The co-clustering matrices (see \cref{fig:coclustering-matrix-country}) show that changes after 2020 are significantly more pronounced, with an unclear block structure evident until 2021, after which the situation stabilizes, a pattern that is less pronounced when aggregated at the regional level (see \cref{fig:coclustering-matrices}). 

\newpage
\begin{table}[h]
\centering
\caption{List of all included countries and cultural regions in the analysis. Note that the listing of individual countries in the list of cultural regions indicates which countries we assign to which cultural region, and not which countries are actually included in the final aggregation.} \label{tab:list-of-regions-and-countries}
\begin{footnotesize}
\begin{tabular}{p{14cm}}
\hline
\textbf{List of countries ($N=112$)} \\ \hline
Afghanistan, Albania, Algeria, Argentina, Australia, Austria, Bangladesh, Barbados, Belgium, Bolivia, Bosnia and Herzegovina, Brazil, British Indian Ocean Territory, Bulgaria, Cambodia, Cameroon, Canada, Cayman Islands, Chile, China, Colombia, Congo, Costa Rica, Croatia, Cuba, Cyprus, Democratic Republic of the Congo, Denmark, Ecuador, Egypt, El Salvador, Estonia, Ethiopia, Finland, France, Gabon, Germany, Ghana, Greece, Greenland, Guatemala, Guernsey, Haiti, Honduras, Hungary, Iceland, India, Indonesia, Iran, Iraq, Ireland, Israel, Italy, Ivory Coast, Jamaica, Japan, Kazakhstan, Kenya, Kuwait, Laos, Malaysia, Malta, Mexico, Moldova, Mongolia, Morocco, Myanmar, Nepal, Netherlands, New Zealand, Nigeria, North Korea, Norway, Pakistan, Panama, Peru, Philippines, Poland, Portugal, Qatar, Romania, Russia, Rwanda, Sao Tome and Principe, Saudi Arabia, Senegal, Serbia, Singapore, Slovakia, Somalia, South Africa, South Korea, Spain, Sri Lanka, Sudan, Sweden, Switzerland, Syria, Taiwan, Tanzania, Thailand, Tunisia, Turkey, Uganda, Ukraine, United Arab Emirates, United Kingdom, United States of America, Vatican City, Venezuela, Vietnam, Zimbabwe \\ \hline
\textbf{List of cultural regions ($N=11$)} \\ \hline
\textbf{Anglo America} (United States of America, Canada), \textbf{Australia/Oceania} (Australia, New Zealand, Fiji, Kiribati, Marshall Islands, Micronesia, Nauru, Palau, Papua New Guinea, Samoa, Solomon Islands, Tonga, Tuvalu, Vanuatu, Cook Islands, Niue, Tokelau), \textbf{Western Europe} (Austria, Belgium, France, Germany, Ireland, Italy, Luxembourg, Netherlands, Portugal, Spain, Switzerland, United Kingdom, Andorra, Monaco, Liechtenstein, Malta, Iceland, Denmark, Norway, Finland, F{\o}royar, Guernsey, Isle of Man, Jersey, San Marino, Gibraltar, Greenland, Akrotiri and Dhekelia, Cyprus, Greece, Sweden, Vatican City), \textbf{Eastern Europe} (Albania, Bosnia and Herzegovina, Bulgaria, Croatia, Czech Republic, Estonia, Hungary, Kosovo, Latvia, Lithuania, Moldova, Montenegro, North Macedonia, Poland, Romania, Serbia, Slovakia, Slovenia, Ukraine, Georgia, Armenia, Azerbaijan, Turkey), \textbf{Russia} (Russia, Kazakhstan, Kyrgyzstan, Tajikistan, Uzbekistan, Turkmenistan, Belarus), \textbf{Latin America} (Argentina, Bolivia, Brazil, Chile, Colombia, Costa Rica,Cuba, Dominican Republic, Ecuador, El Salvador, Guatemala, Honduras, Mexico, Nicaragua, Panama, Paraguay, Peru, Uruguay, Venezuela, Belize, Haiti, Guyana, Suriname, Trinidad and Tobago, Saint Lucia, Saint Vincent and the Grenadines, Saint Kitts and Nevis, Anguilla, Antigua and Barbuda, Barbados, Bermuda, British Virgin Islands, Cayman Islands, Dominica, Falkland Islands, Grenada, Jamaica, Montserrat, South Georgia and the South Sandwich Islands, Turks and Caicos Islands), \textbf{Orient} (Algeria, Bahrain, Egypt, Iran, Iraq, Israel, Jordan, Kuwait, Lebanon, Libya, Morocco, Oman, Qatar, Saudi Arabia, Sahrawi Arab Democratic Republic, Sudan, Syria, Tunisia, United Arab Emirates, Yemen, Afghanistan), \textbf{Sub-Saharan Africa} (Angola, Benin, Botswana, Burkina Faso, Burundi, Cameroon, Cape Verde, Central African Republic, Chad, Comoros, Congo, Democratic Republic of the Congo, Equatorial Guinea, Eritrea, Eswatini, Ethiopia, Gabon, Gambia, Ghana, Guinea, Guin{\'e}-Bissau, Kenya, Lesotho, Liberia, Madagascar, Malawi, Mali, Mauritania, Mauritius, Mozambique, Namibia, Niger, Nigeria, Rwanda, Sao Tome and Principe, Senegal, Seychelles, Sierra Leone, South Africa, South Sudan, Tanzania, Togo, Uganda, Zambia, Zimbabwe, Djibouti, Somalia, Ivory Coast, Saint Helena, Ascension and Tristan da Cunha), \textbf{East Asia} (China, Japan, North Korea, South Korea, Taiwan, Mongolia), \textbf{South Asia} (Bangladesh, Bhutan, India, Maldives, Nepal, Pakistan, Sri Lanka, British Indian Ocean Territory), \textbf{Southeast Asia} (Brunei, Cambodia, Indonesia, Laos, Malaysia, Myanmar, Philippines, Singapore, Thailand, Timor-Leste, Vietnam)  \\ \hline
\end{tabular}
\end{footnotesize}
\end{table}

\begin{table}[h]
\centering
\begin{footnotesize}
\caption{Detailed Topic Description. We decided to use k = 15 topics for our analysis.} \label{tab:topics}
\begin{tabular}{p{0.8cm}p{4cm}p{4cm}p{4cm}}
\hline
\textbf{Topic} & \textbf{k = 20} & \textbf{k = 15} &  \textbf{k = 10} \\ \hline
\textbf{ 0}  & \textbf{Law:} court, case, law, lawyer, judge, justice, decision, legal, claim, trial, right, order, lawsuit, charge, statement                                        & \textbf{Law:} court, case, law, lawyer, judge, justice, decision, legal, claim, trial, charge, federal, order, statement, right                                      & \textbf{Law:} court, case, law, lawyer, judge, justice, decision, trial, legal, charge, claim, federal, order, sentence, statement                                      \\
\textbf{ 1}  & \textbf{Politics:} party, election, liberal, conservative, vote, leader, government, campaign, candidate, minister, ndp, premier, political, trudeau, poll             & \textbf{Politics:} government, minister, party, liberal, leader, election, trudeau, conservative, federal, vote, province, prime, premier, ottawa, political         & \textbf{Politics:} government, minister, party, liberal, leader, trudeau, federal, election, conservative, province, vote, support, prime, public, premier              \\
\textbf{ 2}  & \textbf{Public Health/Healthcare:} health, care, patient, hospital, medical, child, doctor, drug, mental, treatment, study, service, system, support, emergency        & \textbf{Public Health/Healthcare:} health, care, patient, hospital, child, medical, doctor, study, drug, system, service, support, mental, worker, treatment         & \textbf{Public Health/Healthcare:} health, covid-19, case, care, province, vaccine, public, patient, hospital, pandemic, test, number, death, ontario, home             \\
\textbf{ 3}  & \textbf{Conflicts:} ukraine, war, military, russia, israel, russian, attack, force, country, kill, official, state, strike, security, president                        & \textbf{Conflicts:} country, military, war, ukraine, russia, israel, attack, force, russian, official, kill, security, government, state, strike                     & \textbf{Conflicts (internal and external):} school, student, child, war, military, ukraine, country, israel, russia, force, attack, russian, kill, university, parent   \\
\textbf{ 4}  & \textbf{Corporate:} company, technology, business, product, data, global, industry, customer, platform, solution, system, lead, research, experience, create           & \textbf{Corporate:} company, statement, information, million, result, business, technology, financial, security, product, release, management, risk, project, future & \textbf{Corporate:} company, statement, business, information, million, result, financial, technology, product, project, release, security, management, future, risk    \\
\textbf{ 5}  & \textbf{Culture:} child, community, book, young, school, kid, part, good, black, home, film, music, kind, long, great                                                  & \textbf{Culture:} child, home, book, community, good, part, young, event, kid, film, black, long, school, experience, great                                          & \textbf{Sport:} game, team, season, point, player, play, goal, score, lead, win, Toronto, good, sport, coach, run                                                       \\
\textbf{ 6}  & \textbf{Economy:} price, market, rate, bank, high, oil, increase, stock, trade, sale, average, growth, economy, investor, inflation                                    & \textbf{Economy:} price, market, rate, bank, high, increase, oil, trade, cost, stock, billion, sale, company, economy, million                                       & \textbf{Economy:} price, market, rate, bank, high, increase, cost, oil, money, food, pay, stock, million, business, trade                                               \\
\textbf{ 7}  & \textbf{Geopolitics:} state, island, south, unite, democratic, french, territory, north, mexico, vote, american, china, british, carolina, france                      & \textbf{American Politics:} trump, state, president, island, china, biden, unite, american, republican, democratic, house, election, vote, south, donald             & \textbf{American Politics:} trump, state, president, island, china, unite, country, biden, american, election, house, democratic, republican, vote, south               \\
\textbf{ 8}  & \textbf{Municipal Government:} housing, council, community, mayor, tax, plan, home, resident, cost, property, pay, building, money, meeting, public                    & \textbf{Crime/ Law Enforcement:} police, officer, charge, arrest, rcmp, investigation, incident, suspect, victim, release, crime, death, shooting, assault, murder   & \textbf{Crime/ Law Enforcement:} police, officer, charge, vehicle, arrest, rcmp, investigation, incident, suspect, release, driver, crime, victim, street, report       \\
\textbf{ 9}  & \textbf{Natural Disaster:} fire, water, weather, river, lake, high, home, storm, wind, snow, damage, resident, temperature, expect, crew                               & \textbf{Natural Disaster:} fire, water, climate, river, weather, lake, high, storm, home, wind, expect, temperature, snow, damage, resident                          & \textbf{Infrastructure:} fire, water, community, park, home, building, road, resident, project, close, street, open, part, river, highway   \\ \hline                            
\end{tabular}
\end{footnotesize}
\end{table}

\begin{table}[h]
\centering
\begin{footnotesize}
\addtocounter{table}{-1}
\caption{Detailed Topic Description. We decided to use k = 15 topics for our analysis.} 
\begin{tabular}{p{0.8cm}p{4cm}p{4cm}p{4cm}}
\hline
\textbf{Topic} & \textbf{k = 20} & \textbf{k = 15} &  \textbf{k = 10} \\ \hline
\textbf{ 10} & \textbf{Pandemic:} case, covid-19, school, health, vaccine, student, province, public, test, death, number, report, virus, coronavirus, ontario                        & \textbf{Pandemic:} covid-19, case, school, health, student, vaccine, province, public, test, number, death, virus, report, coronavirus, ontario                      & -                                                                                                                                                              \\
\textbf{ 11} & \textbf{Sport:} game, team, season, point, player, score, goal, play, win, lead, coach, league, sport, run, toronto                                                    & \textbf{Sport:} game, team, season, point, player, score, goal, play, win, lead, coach, league, sport, run, Toronto                                                  & -                                                                                                                                                              \\
\textbf{ 12} & \textbf{Local Businesses:} open, business, home, space, restaurant, centre, park, owner, community, shop, pandemic, food, close, local, room                           & \textbf{Development:} community, business, housing, home, building, program, plan, council, open, project, park, space, support, centre, million                     & -                                                                                                                                                              \\
\textbf{ 13} & \textbf{Mobility:} vehicle, road, car, driver, highway, street, close, crash, bus, Toronto, truck, driving, traffic, transit, drive                              & \textbf{Mobility:} vehicle, road, car, driver, highway, street, close, bus, crash, truck, toronto, driving, traffic, transit, drive                            & -                                                                                                                                                              \\
\textbf{ 14} & \textbf{Travel:} woman, flight, olympic, air, man, airport, travel, plane, international, passenger, vancouver, sport, shelter, fly, female                            & \textbf{Travel:} woman, flight, travel, air, airport, international, olympic, man, plane, vancouver, passenger, board, country, fly, sport                           & -                                                                                                                                                              \\
\textbf{ 15} & \textbf{Crime/ Law Enforcement:} police, officer, charge, arrest, rcmp, investigation, incident, suspect, victim, release, crime, death, assault, shooting, murder     & -                                                                                                                                                           & -                                                                                                                                                              \\
\textbf{ 16} & \textbf{Public Policy:} government, province, federal, worker, program, project, plan, million, nation, alberta, ontario, minister, union, indigenous, support         & -                                                                                                                                                           & -                                                                                                                                                              \\
\textbf{ 17} & \textbf{Corporate:} company, statement, million, information, result, financial, release, security, risk, mine, net, cash, capital, management, project                & -                                                                                                                                                           & -                                                                                                                                                              \\
\textbf{ 18} & \textbf{Politics:} minister, government, country, china, prime, trudeau, chinese, protest, foreign, national, state, leader, security, official, political             & -                                                                                                                                                           & -                                                                                                                                                              \\
\textbf{ 19} & \textbf{American Politics:} trump, president, biden, house, republican, white, donald, state, american, administration, democrat, washington, border, senate, campaign & -                                                                                                                                                           & -                                                                                                                                                              \\ \hline
\end{tabular}
\end{footnotesize}
\end{table}

\appendix

\end{document}